\documentclass[preprint2]{aastex}

\newcommand{\beq}{\begin{equation}}
\newcommand{\eeq}{\end{equation}}
\newcommand{\bed}{\begin{displaymath}}
\newcommand{\eed}{\end{displaymath}}
\newcommand{\beqa}{\begin{eqnarray}}
\newcommand{\eeqa}{\end{eqnarray}}
\newcommand{\beqan}{\begin{eqnarray*}}
\newcommand{\eeqan}{\end{eqnarray*}}
\newcommand{\p}{\partial}
\newcommand{\ex}{{\rm exp}}

\newcommand{\fn}{\footnote}
\newcommand{\bfl}{\begin{flushleft}}
\newcommand{\efl}{\end{flushleft}}
\newcommand{\bfr}{\begin{flushright}}
\newcommand{\efr}{\end{flushright}}
\newcommand{\eps}{\epsilon}
\newcommand{\lae}{\mathrel{<\kern-1.0em\lower0.9ex\hbox{$\sim$}}}
\newcommand{\gae}{\mathrel{>\kern-1.0em\lower0.9ex\hbox{$\sim$}}}
\newcommand{\noi}{\noindent}
\newcommand{\btab}{\begin{tabbing}}
\newcommand{\etab}{\end{tabbing}}
\newcommand{\nb}{\nabla}
\newcommand{\bcdot}{{\bf \cdot}}
\newcommand{\non}{\nonumber}
\newcommand{\rcl}{{\rm cl}}
\newcommand{\mx}{{\rm max}}

\newcommand{\bfig}{\begin{figure}}
\newcommand{\efig}{\end{figure}}

\begin{document}
\title{THE STRUCTURE AND EVOLUTION OF MAGNETIZED \\ 
CLOUD CORES IN A ZERO--DENSITY BACKGROUND\altaffilmark{1}}
\author{CHARLES L.\ CURRY\altaffilmark{2}}
\affil{Department of Physics, 
University of Waterloo, Waterloo, ON  N2L 3G1}
\affil{and} 
\affil{Department of Physics and Astronomy, 
University of Western Ontario, London, ON  N6A 3K7}
\author{and}
\author{STEVEN W.\ STAHLER}
\affil{Astronomy Department, University of California, Berkeley, 
CA  94720}
\altaffiltext{1}{\sc To appear in ApJ: June 20, 2001}
\altaffiltext{2}{Email: curry@astro.uwo.ca} 

\begin{abstract} 
Molecular-line observations of star-forming cloud cores indicate that
they are not the flattened structures traditionally considered by
theory. Rather, they are elongated, perhaps in the direction of their
internal magnetic field. We are thus motivated to consider the
structure and evolution of axisymmetric, magnetized clouds that start from
a variety of initial states, both flattened (oblate) and elongated
(prolate). In this first contribution, the clouds are of fixed mass, 
and are surrounded by a fictitious medium of zero density and finite
pressure. 
We devise a new technique, dubbed the $q$-method, that allows
us to construct magnetostatic equilibria of any specified shape. The
mass loading of the field lines then follows from the self-consistent
model solution, just the reverse of the standard procedure. 
We find, in agreement with previous authors, that the field lines in 
oblate clouds bend inward. However, those in prolate clouds bow outward, 
confining the structures through magnetic tension.

We next follow the quasi-static evolution of these clouds via ambipolar
diffusion. An oblate cloud either relaxes to a magnetically force-free
sphere or, if sufficiently massive, flattens along its polar axis
as its central density runs away. A prolate cloud always relaxes to a
sphere of modest central density. We finally consider the evolution of
an initially spherical cloud subject to the tidal gravity of neighboring
bodies. Although the structure constricts equatorially, it also
shortens along the pole, so that it ultimately flattens on the way to
collapse. In summary, none of our initial states can evolve to the point
of collapse while maintaining an elongated shape. We speculate that this
situation will change once we allow the cloud to gain mass from its
environment.
\end{abstract}
\keywords{ISM: clouds --- ISM: structure --- ISM: magnetic fields 
--- MHD --- stars: formation}

\section{INTRODUCTION}
\label{sec-intro}

After two decades of intense observational effort, astronomers have 
learned much about the nature of the dense, molecular cloud cores that 
form low-mass stars (see, e.g., Myers 1999).  A typical core, with a mass 
of several solar masses and a diameter of order 0.1 pc, has a central 
density somewhat in excess of $10^4$ cm$^{-3}$ and a kinetic temperature 
of about 10 K.  Zeeman measurements, still only marginally feasible at 
such densities, suggest magnetic field strengths no larger than 30 $\mu$G 
(Crutcher 1999). These 
figures refer specifically to dense cores in relatively sparse environments 
like Taurus-Auriga. Those within regions such as Orion, that form high-mass 
stars, appear to be more massive and warmer, although the measurements here 
are generally less precise because of the greater distances 
involved (e.g., Jijina, Myers, \& Adams 1999).

To see how a dense core evolves to form a star, one must first have a 
physical understanding of its structure. A core's internal velocity 
dispersion, as gauged from the linewidths of tracer molecules like NH$_3$ 
and CS, depends on its mass.  In low-mass cores, the dispersion is typically 
0.2--0.4 km s$^{-1}$, significantly lower than the background gas. 
This range in dispersion matches that expected from the virial theorem for 
objects of the appropriate mass and size. The line profiles, moreover, show 
only slight nonthermal broadening (Myers \& Benson 1983).  Finally, the 
internal magnetic field strength is also consistent with the virial 
expectation, and with approximate equipartition of thermal and magnetic 
energies (Myers \& Goodman 1988). These facts together imply that the 
typical low--mass core is in dynamical equilibrium, 
supported against self-gravity by a combination of ordinary gas pressure 
and the Lorentz force associated with a largely static magnetic field 
(McKee et al.\ 1993).

Theorists have long recognized that such a magnetostatic structure evolves
quasi-statically through ambipolar diffusion, i.e., the relative drift of
neutrals and ions (Mestel \& Spitzer 1956). As the configuration slips through
the ambient magnetic field, its central density gradually increases to the
point where rapid, protostellar collapse begins (Nakano 1979).  While this
basic picture continues to provide a framework for our conception of dense
core evolution, its quantitative implementation is not without problems. 
One important issue, and the focus of this paper, concerns the 
three-dimensional {\it shape} of the objects.

Radio maps in a number of tracer lines show dense cores to be distinctly
nonspherical, whether they are in loose associations (Benson \& Myers 
1989) or massive, turbulent complexes (Harju, Walmsley, \& Wouterloot 
1993), and whether they contain stars or not (Jijina et al.\ 1999). 
The observed distribution of projected axial ratios is well matched 
if the cores are intrinsically prolate structures, with a random 
orientation of their long axes in the plane of the sky (Myers et al.\
1991; Ryden 1996). Theoretical models of magnetostatic clouds, on the other 
hand, generally predict that they are {\it oblate}, flattened along the 
direction of the background magnetic field (Mouschovias 1976b; Tomisaka, 
Ikeuchi, \& Nakamura 1988a).  Under ambipolar diffusion, the central 
region flattens even more as the central density climbs (Tomisaka, Ikeuchi, 
\& Nakamura 1990). 

A useful step toward reconciling theory and observation is to broaden the
range of models and consider cores with a {\it variety} of intrinsic shapes,
including prolate configurations. This is the first task of the present 
study. The essential reason why previous authors obtained flattened structures
is not difficult to see. In solving the equations for force balance, one
customarily specifies the amount of mass loading each magnetic flux tube
(Mouschovias 1976a). Lacking any observational indication of this 
mass-to-flux distribution, theorists have taken it from geometrically 
simple reference
states, such as uniform spheres or cylinders threaded by a spatially 
constant field. Such configurations, however, are clearly {\it not} in
force balance. If allowed to evolve, they immediately collapse along the field
until dynamical equilibrium is restored (see, e.g, Fig.\ 2 of Fiedler \&
Mouschovias 1993). The magnetostatic structures resulting from such collapse
are inevitably flattened, and become more so during the subsequent phase of
quasi-static settling via ambipolar diffusion.  This qualitative behavior is 
unaltered by the presence of isotropic, subsonic turbulence 
(see, e.g., the schematic treatment of Lizano \& Shu 1989). 

To remedy this situation, we break from tradition and use the core shape
itself as input, rather than the unknown mass-to-flux distribution. That is,
we solve the magnetostatic equations subject to the constraint that the gas
pressure fall to some uniform, ambient value along a specified boundary. The
mass-to-flux distribution then results from the self-consistent numerical
solution. We use this technique, dubbed the {\it q-method}, to construct a
variety of equilibria, with shapes ranging from oblate to prolate. We 
further distinguish two classes of states, those which are spatially
isolated, and those subject to the tidal gravitational field of nearby 
cores. Study of the first class allows ready comparison with previous 
authors, while the tidal boundary condition approximates the observed 
situation within filamentary clouds, where neighboring cores are 
typically separated by only 0.5--1.0~pc (e.g., L1495 in Taurus; Onishi 
et al.\ 1996).

Our next task is to follow the quasi-static evolution of selected initial
states through ambipolar diffusion. In this paper, we assume each core
to be of fixed total mass. That is, we take the ambient pressure to arise
from a fictitious medium of negligible mass density and correspondingly high
temperature. Our numerical results first confirm that isolated, oblate 
structures undergo significant central flattening as their density rises 
steeply.  Isolated, prolate structures experience no runaway increase in 
density, but instead relax to spheres. Finally, an initially spherical, 
but tidally stressed configuration does undergo central condensation, but 
concurrently shrinks 
along the polar axis. In summary, none of our states both evolves to a high
central density and maintains an elongated shape, as the observations seem to 
demand. 

Section 2 below details our numerical method for constructing magnetostatic
equilibria with specified shapes. In Section 3, we describe the physical
properties of these equilibria. Section 4 then presents numerical results 
for the quasi-static evolution of a few representative models.  Finally, 
Section 5 assesses our results in light of previous observational and 
theoretical work.  We conclude by emphasizing the potential value in 
relaxing the assumption of fixed core mass.  In a second paper, we
shall explore cloud evolution under these more general circumstances. 

\section{FORMULATION OF THE PROBLEM}
\label{sec-form}

\subsection{Governing Equations of Equilibrium}

The theory of self-gravitating, magnetized cloud equilibria, first 
discussed by Mestel (1965) and Strittmatter (1966), has been 
extended and reformulated by many authors---notably Mouschovias 
(1976a), Nakano (1979, 1984), and Tomisaka et al.\ (1988a).  Here, 
we briefly review the basic equations, before passing to a discussion 
of the features unique to our method. 

The governing equations are the condition of force balance, Poisson's 
equation, and Amp\`ere's law:
\beqa
-\nb P - \rho \nb \psi + \frac{1}{c}~{\bf j} \times 
{\bf B} &=& 0, \label{eq-fbal}  \\
\nb^2 \psi &=& 4\pi G \rho, \label{eq-poisson} \\
\nb \times {\bf B} &=& \frac{4\pi}{c} {\bf j} \label{eq-ampere}, 
\eeqa
where the chosen symbols are standard.  Our adopted equation of state 
is that of an isothermal gas, $P = a^2 \rho$, where $a$ is the constant
sound speed.  The ancillary relation
\beq
{\bf B} = \nb \times {\bf A}, \label{eq-magp}
\eeq
where ${\bf A}$ is the magnetic vector potential, ensures that 
$\nb \bcdot {\bf B} = 0$ is satisfied identically. 

We adopt a cylindrical coordinate system $(r,\phi,z)$ whose origin lies 
at the cloud center; see Figure 1.  We assume axial
symmetry about the $z$-axis, and reflection symmetry about the plane 
$z = 0$.  Hence, only the quadrant $r \geq 0,~z \geq 0$ needs 
to be considered explicitly.  We assume that ${\bf B}$ is poloidal, 
and hence ${\bf A}$ toroidal. We thus take ${\bf A} = 
A(r,z) ~{\bf \hat{e}}_{\phi}$. Alternatively, ${\bf B} = -r^{-1} 
{\bf \hat{e}}_{\phi} 
\times \nb \Phi$, where $\Phi$ is the scalar function defined by
\bed
\Phi (r,z) \equiv r A (r,z). 
\eed
The quantity $\Phi$ is invariant along each magnetic surface: ${\bf B} 
\bcdot \nb \Phi = 0$.  Since it is proportional to the usual 
magnetic flux $\Phi_B$, i.e.\ $\Phi = \Phi_B/2\pi$, we refer to $\Phi$ 
loosely as the ``magnetic flux,'' and use it to label field lines.

It is convenient to resolve the forces in equation (\ref{eq-fbal}) both 
along and across field lines.  Following Dungey (1953) and Mouschovias 
(1976a), we introduce the scalar function
\beq
q \equiv P~\ex (\psi/a^2) = a^2 \rho ~\ex (\psi/a^2). 
\label{eq-qdef}
\eeq
Equation (\ref{eq-fbal}) may be rewritten in terms of $\Phi$ and $q$ as 
\beq
j ~\nb \Phi = c r ~\ex (-\psi/a^2) ~\nb q, 
\label{eq-fbalq1}
\eeq
where we have noted that ${\bf j} = j ~{\bf \hat{e}}_{\phi}$. 
While $q$ depends on both $r$ and $z$, its real utility comes 
from the fact that it is a function of $\Phi$ alone.  That is, 
it follows from equation (\ref{eq-fbalq1}) and ${\bf B} \bcdot 
\nb \Phi = 0$ that ${\bf B} \bcdot \nb q = 0$.  This property 
represents force balance in the direction parallel to ${\bf B}$.
Since $q = q(\Phi)$, we may use equation (\ref{eq-fbalq1}) to 
express force balance perpendicular to the field as 
\beq
\frac{j}{cr} \ex (\psi/a^2) = \frac{dq (\Phi)}{d\Phi}.
\label{eq-fbalq2}
\eeq
Using equations (\ref{eq-magp}) and (\ref{eq-fbalq2}), equation 
(\ref{eq-ampere}) becomes 
\beqa
\frac{\p}{\p r}\left[\frac{1}{r}\frac{\p}{\p r} (rA)\right] + 
\frac{\p^2 A}{\p z^2} &=& \non \\
-4\pi r ~\ex (-\psi/a^2)~ 
\frac{dq (\Phi)}{d\Phi} && {\rm (interior)} \label{eq-aeq} \\
0 && {\rm (exterior)},
\non
\eeqa
where the terms ``interior'' and ``exterior'' refer to the regions
inside and outside the cloud, respectively.  As in previous studies, 
the cloud exterior is assumed to be composed of a hot and tenuous gas 
that has no dynamical effect except to exert a finite pressure, $P_0$, 
on the cloud boundary.  
The exterior region is itself bounded by a cylinder of radius $R$ 
and half-height $Z$ (see Figure 1). It is on this cylindrical surface 
that we apply our boundary conditions (see \S \ref{sec-bcs} below).

Equation (\ref{eq-aeq}) is to be solved throughout the entire volume, 
simultaneously with Poisson's equation (\ref{eq-poisson}), 
\beqa
\frac{1}{r}\frac{\p}{\p r}\left(r\frac{\p \psi}{\p r}\right) +
\frac{\p^2 \psi}{\p z^2} &=& 
\non \\
\frac{4\pi G}{a^2} q(\Phi) ~\ex (-\psi/a^2) && {\rm (interior)} 
\label{eq-psieq} \\
0 && {\rm (exterior)},
\non
\eeqa
where we have made use of equation (\ref{eq-qdef}) on the right-hand 
side of the interior equation.  
Solving the equations in this form guarantees that the relevant continuity
conditions for the gravitational and magnetic fields across the cloud
surface are satisfied (Mouschovias 1976a).

\subsection{Boundary Conditions}
\label{sec-bcs}

The boundary conditions are applied along the cylindrical surface 
bounding the exterior zero-density region.
The boundary condition on $A$ is that the magnetic field approach 
a uniform background value, $B = B_{\infty}$, at the boundary.  
That is,   
\beqa
A &=& \frac{B_{\infty} r}{2}~~~~ {\rm at} ~~~ z = Z, \non \\ 
A &=& \frac{B_{\infty} R}{2}~~~~ {\rm at} ~~ r = R. 
\label{eq-abc}
\eeqa
We consider two types of boundary condition on the potential $\psi$, 
reflecting different assumptions regarding the external medium. 
\vskip 0.1cm
\noi
{\it (i) Point-Mass, or ``Isolated'' Boundary Condition.} 
This boundary condition has been adopted in most previous studies.  
It asserts that $\psi$ in the exterior region far from the cloud is 
indistinguishable from that of a point particle of the same mass: 
\beq
\psi (r,z) = \frac{-GM}{(r^2 + z^2)^{1/2}}~~~~~~~{\rm at}~~ 
r = R, ~z = Z, 
\label{eq-ptbc}
\eeq
where $M$ is the mass of the cloud.  In practice, we have used $R = 2R_0$ 
and $Z = 2Z_0$ when applying this boundary condition. Here $R_0$ and $Z_0$ 
are, respectively, the radial and vertical extent of the cloud. 
Calculations employing larger cylinders give essentially identical results.
\vskip 0.1cm
\noi
{\it (ii) Periodic, or ``Tidal'', Boundary Condition.} 
This boundary condition dictates that the gravitational force vanish 
along some surface at a finite distance from the cloud, as is appropriate 
when surrounding matter is present. We take these surfaces to be the top 
and bottom of our cylindrical outer boundary (e.g.\ Fiedler \& Mouschovias 
1992).  The conditions on the potential then become
\beqa
\frac{\p \psi}{\p z} &=& 0 ~~~~{\rm at}~~~z = Z, \non \\
\psi &=& {\rm constant} ~~~~{\rm at}~~~r = R. \label{eq-tidal}
\eeqa
Note that the first condition holds if the cloud of interest is
one of an infinite chain, with a spacing of $2Z$. The constancy
of the potential at $r=R$ simulates the presence of a larger background
filament.\fn{Our periodic boundary condition differs from that used by 
Lizano \& Shu (1989), in that the latter authors took the gravitational 
potential to be constant on a specified locus corresponding to a chain 
of point masses. Moreover, they forced ${\bf B}$ to equal a uniform 
background value on that surface.}  

Finally, the assumed reflection symmetry about $z = 0$ implies that 
\bed
\frac{\p \psi}{\p z} = 0, ~~~~ \frac{\p A}{\p z} = 0,~~~~~~{\rm at}~~~~ 
z=0,
\eed
while the vanishing of the radial gravitational force and the lack of 
magnetic sources along the $z$-axis imply
\bed
\frac{\p \psi}{\p r} = 0, ~~~~ A (0,z) = 0 ~~~~ {\rm at} ~~~~ r=0.
\eed

\subsection{Construction of Initial States: The $q-$Method}
\label{sec-constr}

Consider a single magnetic flux tube penetrating the cloud, containing 
mass $\delta m$ and flux $\delta \Phi$.  The former is given by 
(Mouschovias 1976a) 
\bed
\delta m (\Phi) = 2 \int_0^{Z_\rcl (r)} dz \int_{r(\Phi, z)}^
{r(\Phi + \delta \Phi, z)} 2\pi r \rho (r,z) dr, 
\eed
where $Z_\rcl (r)$ describes the boundary shape (Figure 1).  Changing 
variables from $r$ to $\Phi$, we have $dr = d\Phi (\p r/\p \Phi)$, so 
that a trivial integration over $\Phi$ yields the cloud's mass-to-flux 
distribution
\beqa
\frac{dm}{d\Phi} &=& \frac{4\pi}{a^2}~ q(\Phi) \int_0^{Z_\rcl (\Phi)} 
r(\Phi, z) \frac{\p r(\Phi, z)}{\p \Phi} \non \\ 
&\times& \ex \left[-\frac{\psi (\Phi, z)}{a^2}\right] ~dz, 
\label{eq-dmdphi} 
\eeqa
where we have substituted for $\rho$ from equation (\ref{eq-qdef}). 
Here $m (\Phi)$ is the total mass contained within a given 
(axisymmetric) surface of constant magnetic flux $\Phi$:
\beq
m (\Phi) = \int_0^{\Phi} \frac{dm (\Phi)}{d\Phi}~d\Phi.
\label{eq-mdef}
\eeq
Denoting the total flux enclosed in the cloud by $\Phi_0 \equiv 
\Phi (R_0)$, the total mass is then $M \equiv m(\Phi_0)$. 

The customary procedure for obtaining magnetostatic equilibria
is to solve equation (\ref{eq-dmdphi}) for $q$, obtaining
\beqa
q (\Phi) &=& \frac{a^2}{4\pi}~ \frac{dm (\Phi)}{d\Phi}\Bigg/ 
\int_0^{Z_\rcl (\Phi)} r(\Phi, z)\frac{\p r}{\p \Phi} \non \\
&\times& \ex 
\left[\frac{-\psi (\Phi, z)}{a^2}\right]~dz. 
\label{eq-tmqdef} 
\eeqa
This expression is then used in the right-hand sides of equations
(\ref{eq-aeq}) and (\ref{eq-psieq}).  Note that all quantities on the 
right-hand side of equation (\ref{eq-tmqdef}) are determined iteratively 
in a numerical scheme, {\it except} $dm/d\Phi$, which is specified 
a priori (\S \ref{sec-intro}).  

As noted in \S \ref{sec-intro}, we invert the customary procedure 
when constructing our initial states.  We do not specify a priori 
$dm/d\Phi$ within the cloud; instead, the latter is a {\it result 
of} the magnetostatic calculation.  Our method is to specify the shape 
of the cloud boundary $Z_\rcl (r)$.
On the boundary, the pressure $P_0$ is a known constant, while the 
potential, $\psi_0 \equiv \psi [Z_\rcl (r)]$, is obtained through
concurrent solution of Poisson's equation. 
From the definition of $q$, equation (\ref{eq-qdef}), this quantity 
is therefore determined {\it everywhere within the cloud} by
\beq
q (\Phi) = P_0 ~ \ex (-\psi_0/a^2), 
\label{eq-qmeth} 
\eeq
where $\psi_0$ is known along each flux tube. 
With successive iterations, estimates of $q, ~\psi$, and $A$ improve, 
until convergence is achieved. (For convergence criteria and other 
numerical details, see the Appendix.)  Henceforth, we refer to this 
technique as the {\it q-method}. We revert to the customary, or {\it 
free-boundary} method when constructing equilibria that have evolved 
from the initial state.

\subsection{Quasi-static Evolution}
\label{sec-quasi}

At typical molecular densities of $10^3-10^5$ cm$^{-3}$, the 
cloud can remain in equilibrium with a frozen field for roughly 
$10^6$ yr. Over longer periods, however, self-gravity causes the 
neutral species to drift inward relative to the ions, which are 
tied to the magnetic field lines (Mestel \& Spitzer 1956).
The drift velocity of neutrals with respect to ions, ${\bf v}_d 
\equiv {\bf v} - {\bf v}_i$ (no subscript indicating neutrals), 
depends both on the cloud's level of ionization and on the amount 
of collisional drag between neutrals and ions.  At the low density  
contrasts characterizing our equilibria, $|{\bf v}| \ll a$, 
so that the $\rho~D{\bf v}/Dt$ term normally appearing on the right-hand 
side of the equation of motion (equation \ref{eq-fbal}) can safely be 
ignored.  We are thus describing the {\it quasi-static} phase of 
evolution (Nakano 1984), during which the cloud slowly progresses 
along a sequence of exact, magnetostatic equilibria. 
As the cloud evolves, it becomes more centrally condensed, and 
gravity becomes relatively more important than magnetic forces 
near the cloud center.  Once the neutral velocities there approach 
the sound speed, the quasi-static approximation breaks down, and 
the cloud undergoes essentially hydrodynamic collapse (e.g., 
Fiedler \& Mouschovias 1993).  In this paper, we shall follow the 
quasi-static phase only, while also keeping track of the neutral 
velocities during the evolution.  

The drift speed is given by 
\beqa
{\bf v}_d \equiv {\bf v} - {\bf v}_i = -\frac{\tau_i}{\rho_i} 
(\nb \times {\bf B}) \times {\bf B} \non \\
= -(\gamma {\cal C})^{-1} \rho^{-3/2}
\frac{dq}{d\Phi}~\ex (-\psi/a^2) \nb \Phi,
\label{eq-vd}
\eeqa
where we have chosen the opposite sign convention to Nakano (1979), 
who gave an equivalent expression for the magnetic field (or ion) drift.
Here $\tau_i = (\gamma \rho)^{-1}$ is the damping time of ions 
relative to neutrals, $\rho_i$ is the mass density of the ions, 
and $\gamma = 4.28 \times 10^{13}$ cm$^3$ g$^{-1}$ s$^{-1}$ is the 
frictional drag coefficient (Nakano 1979).  In the second step, 
we have made use of equations (\ref{eq-ampere}) and (\ref{eq-fbalq2}), 
and have assumed $\rho_i = {\cal C} \rho^{1/2}$, where ${\cal C} = 
4.46 \times 10^{-16}$ g$^{1/2}$ cm$^{-3/2}$ (Nakano 1979; Elmegreen 
1979).  This expression for $\rho_i$ assumes that ionization from 
cosmic rays balances recombination. 
We ignore negatively charged grains, since their effect is  
small for the typical neutral gas densities we consider
in this paper ($n \ll 10^7$ cm$^{-3}$).  For a poloidal magnetic 
field that decreases outward in the cloud, both $dq/d\Phi$ and 
$\p \Phi/\p r$ are positive, so $v_{d,r} < 0$, indicating inward 
drift of the neutrals (note that $|v_{d,z}| \ll |v_{d,r}|$ during  
most of the quasi-static phase).  
The neutral velocity, ${\bf v}$, 
is calculated using the prescription of Lizano \& Shu (1989).\fn{Note, 
however, that Lizano \& Shu (1989) used values of $\gamma$ and ${\cal C}$ 
which result in drift velocities 80 percent larger than those found in 
Tomisaka et al.\ (1990) and the present paper. Their evolutionary 
timescales are correspondingly shorter.}
The qualitative behavior of both ${\bf v}_d$ and ${\bf v}$ is 
shown in Figure 2.

Neutral-ion drift leads to a redistribution of mass 
with magnetic flux in the cloud.  More precisely, the time rate of 
change of the mass contained within a given (axisymmetric) surface 
of constant magnetic flux $\Phi$ is
\beq
\frac{\p m (\Phi, t)}{\p t} = -\int_{\Phi} \rho~ {\bf v}_d \bcdot {\bf dS},
\label{eq-dmdt1} 
\eeq
where the surface integration is performed along the entire
flux tube $\Phi =$ constant (Figure 2), and where the definition 
(\ref{eq-mdef}) of the mass function has been extended to the 
quasi-static case, i.e. $m = m(\Phi,t)$.  Since ${\bf dS}$ points 
outward by definition, generally opposite to the direction of 
${\bf v}_d$, $\p m/\p t$ is usually positive for all $\Phi < \Phi_0$.  
It then follows that $\Phi_0$ is usually a {\it decreasing} function 
of time, i.e., the cloud as a whole loses flux.  In the present 
coordinates, equations (\ref{eq-vd}) and (\ref{eq-dmdt1}) combine 
to give 
\beqa
\frac{\p}{\p t}~[m(\Phi,t)] &=& \frac{4\pi}{\gamma {\cal C}}
\frac{dq}{d\Phi} \int_0^{Z_\rcl (\Phi)} \rho^{-1/2} \ex 
(-\psi/a^2) \non \\
&\times& r\frac{\p \Phi}{\p r}\left[1 + \left(\frac{\p r}
{\p z}\right)_{\Phi}^2 \right] dz. 
\label{eq-dmdt2} 
\eeqa
The quantity $\p m/\p t$ vanishes at both $\Phi = 0$ and $\Phi = \Phi_0$. 
At each time step $\Delta t$, one may adjust the mass-to-flux distribution 
as follows: 
\beqa
\frac{\p m}{\p \Phi} (\Phi,t+\Delta t) &=& \frac{\p m}{\p \Phi} (\Phi,t) 
\non \\
&+& \Delta t \frac{\p}{\p \Phi} \left[\frac{\p m}{\p t} (\Phi, t)\right].
\label{eq-dmdphi2} 
\eeqa
Given $dm/d\Phi$ at the next time step, $q(\Phi,t+\Delta t)$ 
may be calculated from equation (\ref{eq-tmqdef}).  Recall that 
$(dm/d\Phi)_{t=0}$ is known from the initial state, constructed 
via the $q-$method (equation \ref{eq-dmdphi}).  During the subsequent 
evolution 
{\it the boundary shape is allowed to vary}, with the solution at all 
times obeying the same boundary conditions as in the magnetostatic case. 
Equation (\ref{eq-tmqdef}) is therefore used to calculate $q(\Phi)$
during the quasi-static phase of the calculation, and the new equilibrium
is constructed via the free-boundary method.  Here, the boundary location
$Z_\rcl (\Phi)$ is calculated by finding the locus of points where 
$\rho (\Phi, z) = \rho_0 = P_0/a^2$. 

\subsection{Nondimensionalization and Free Parameters}
\subsubsection{Magnetostatic equations}
\label{sec-nondim}

We nondimensionalize our equations with respect to the sound speed 
$a$, cloud boundary density $\rho_0$, and background magnetic field 
$B_{\infty}$.  From these, we construct the following dimensionless 
quantities for the remaining variables:
\beqan 
&&r' = r\cdot (4\pi G \rho_0)^{1/2}/a, \\
&& z' = z\cdot (4\pi G \rho_0)^{1/2}/a, \\
&&\psi' = \psi/a^2, \\
&& A' = A\cdot (4\pi G \rho_0)^{1/2}/a B_{\infty}, \\
&&\Phi' = \Phi\cdot (4\pi G \rho_0)/a^2 B_{\infty}, \\
&& q' = q/a^2 \rho_0, \\
&&M' = M\cdot (4\pi G \rho_0)^{3/2}/a^3 \rho_0, \\
&& t' = t\cdot (4\pi G \rho_0)^{1/2}.
\eeqan
Observations suggest that the typical cloud core temperature is $T 
\simeq 10$ K (Jijina et al.\ 1999), while a typical number density  
outside cores is $n_{H_2} \sim 10^3$ cm$^{-3}$ (Nercessian et al.\ 1988). 
Using $a = (k_B T/\mu)^{1/2}$ with $\mu = 2.33~m_H$, one then finds 
that $r' = 1$ corresponds to a dimensional length of $L_0 = 0.11$ pc. 
Similarly, $M' = 1$ corresponds to $M_0 = 0.070~M_{\sun}$, and $t' = 
1$ to $t_0 = 5.5 \times 10^5$ yr.

With these definitions, the dimensionless forms of the fundamental 
equations (\ref{eq-aeq}) and (\ref{eq-psieq}) are 
\beqa
\frac{\p}{\p r'}\left[\frac{1}{r'}\frac{\p}{\p r'} (r'A')\right] + 
\frac{\p^2 A'}{\p z'^2} &=& \non \\
-\frac{r'}{2\alpha}~\ex (-\psi')~ 
\frac{dq'}{d\Phi'} && {\rm (interior)} \label{eq-aeqnd} \\
0 && {\rm (exterior)},
\non \\
\non \\
\frac{1}{r'}\frac{\p}{\p r'}\left(r'\frac{\p \psi'}{\p r'}\right) + 
\frac{\p^2 \psi'}{\p z'^2} &=& \non \\
q' ~\ex (-\psi') && {\rm (interior)} \label{eq-psieqnd} \\
0 && {\rm (exterior)},
\non
\eeqa
The dimensionless parameter $\alpha$ appearing on the right-hand 
side of equation (\ref{eq-aeqnd}) is defined in terms of fiducial 
quantities by
\bed
\alpha \equiv \frac{B_{\infty}^2/8\pi}{a^2 \rho_0}. 
\eed
That is, $\alpha$ is the ratio of magnetic pressure far from the 
cloud to gas pressure at the cloud surface.  

\subsubsection{Quasi-static evolution equations}
The dimensionless forms of equations (\ref{eq-vd}) and (\ref{eq-dmdt2}) 
are 
\beq
{\bf v}_d' = -C_1~
\rho'^{-3/2}~\frac{dq'}{d\Phi'}~\ex (-\psi') ~(\nb \Phi)'
\label{eq-vdnd}
\eeq
and 
\beqa
\frac{\p m'}{\p t'} &=& C_2~
\frac{dq'}{d\Phi'} ~\int_0^{Z_\rcl' (\Phi')} \rho'^{-1/2} ~\ex 
(-\psi') \non \\ 
&\times& r' ~\frac{\p \Phi'}{\p r'}~\left[1 + \left(\frac{\p r'}
{\p z'}\right)_{\Phi'}^2 \right] dz',
\label{eq-dmdt2nd} 
\eeqa
where the numerical coefficients on the right-hand sides of 
equations (\ref{eq-vdnd}) and (\ref{eq-dmdt2nd}) are defined in 
terms of dimensional quantities by $C_1 \equiv (4\pi G)^{1/2}/\gamma 
{\cal C} = 0.0480$ and $C_2 \equiv 4\pi C_1 = 0.603$, respectively. 

The point-mass boundary condition (equation \ref{eq-ptbc}), in 
dimensionless form, becomes
\beq
\psi' = \frac{-M'}{4\pi (r'^2 + z'^2)^{1/2}}~~~~~~~{\rm at}~~ 
r' = R',~ z' = Z'. 
\label{eq-ptbcnd}
\eeq
The remaining boundary conditions are just the primed forms of those 
in \S \ref{sec-bcs}, upon making the replacement $B_{\infty} = 1$.

\subsubsection{Free parameters and functions}
\label{sec-param}
To construct equilibria under the point-mass boundary condition, we 
need to specify one parameter, $\alpha$, and one function, the cloud 
shape, $Z_\rcl (r)$.  Measurements of magnetic 
field strengths in low-mass star-forming regions (mostly upper limits) 
suggest a mean of roughly 20 $\mu$G (Crutcher 1999), corresponding to 
$\alpha = 11$.  Given the uncertainty in the field measurements,  
we have performed calculations for $\alpha = 1$ and $\alpha = 10$.

We constructed both spherical clouds and oblate/prolate ellipsoidal 
clouds with shapes given by 
\beq
Z_\rcl' (r') = \frac{Z_0'}{R_0'}(R_0'^2 - r'^2)^{1/2},
\label{eq-shape}
\eeq
where $R_0'$ and $Z_0' = Z_\rcl' (0)$ are the principal axes of the 
ellipsoid.  In the oblate and prolate cases, the cloud's axial ratio 
$Z_0'/R_0'$ was fixed at 1/2 and 2, respectively.  These figures are  
in accord with the most likely intrinsic mean value derived from 
observations (Myers et al.\ 1991; Ryden 1996).  As we describe below in 
\S \ref{sec-init}, equilibrium sequences were constructed by considering 
a range of sizes $R_0'$. 

Finally, under the periodic boundary condition, we need to specify 
the additional parameter $2Z$, the intercore spacing.  
For the typical spacings in Taurus discussed in 
\S \ref{sec-intro}, i.e.\ $0.50 ~{\rm pc}\lae 2Z \lae 1.0 ~{\rm pc}$, 
we have $2.3 \lae Z' \lae 4.5$.  We adopt the value  
$Z' = 2.9$, corresponding to a dimensional spacing of $2Z = 0.65$ pc.
We also fix the extent of the computational volume in the $r$-direction 
at $R' = 2R_0'$.

\section{INITIAL STATES}
\label{sec-init}
In this section, we present the entire set of equilibria obtained 
using the $q-$method (\S \ref{sec-constr}).  We refer to them as 
``initial'' states, since we shall later be concerned with their 
quasi-static evolution via ambipolar diffusion. 

\subsection{Calculation of Equilibrium Sequences}
\label{sec-eqseq}
For a given cloud shape and boundary condition (point-mass or periodic), 
a sequence of states is obtained, each member of which is 
uniquely specified by its center-to-surface density contrast $\rho_c$. 
(Henceforth, we shall use only nondimensional quantities, omitting the 
primes, unless otherwise specified.)
As $\rho_c$ increases, $R_0$ and $M$ first rise to 
maximum values, $R_{0,\mx}$ and $M_\mx$, respectively.  
Both quantities then decline. For much larger $\rho_c$ than covered
in this study, $R_0$ and $M$ should undergo damped, oscillatory
behavior, as found in the sequence of isothermal spheres 
(Chandrasekhar 1939). 

For the low-$\rho_c$ portion of each sequence, we used the radius $R_0$ 
as an input parameter for the models.  That is, we specified the cloud
shape and then used the $q$-method to find both $\rho_c$ and the internal
structure of each model.  At this stage, convergence was not overly 
sensitive to the initial guesses for the gravitational and magnetic 
fields.  In practice, $\psi (r,z) = 0$ and $A (r,z) = r/2$ proved 
satisfactory initial guesses, where the latter corresponds to a 
uniform magnetic field set to the background value.  

In each sequence, the equatorial and polar radius always peaked well 
before the cloud mass. This feature is familiar from the sequence of 
isothermal, nonmagnetic (Bonnor-Ebert) spheres, for which $R_{0,\mx} 
= 1.822$ and $M_\mx \equiv M_{BE} = 52.66$ in our units.  
In general, one cannot proceed to models with higher $\rho_c$ than
that of $R_{0,\mx}$ by specifying the cloud shape. This is because,
along this portion of the sequence, each radius $R_0 < R_{0,\mx}$
may correspond to more than one $\rho_c-$value and internal structure.   
We therefore modified the procedure, with the technique depending
on the particular sequence (see below).  
Proceeding in the appropriate manner to larger $\rho_c$, one 
eventually surpasses the mass peak of the sequence, $M_\mx$.  
Since this peak signifies, at least roughly, the transition to 
dynamically unstable clouds (\S \ref{sec-stab}), we halted the 
search for equilibria soon thereafter.

\subsection{Isolated Clouds}
\label{sec-isol}
\subsubsection{Force-Free States}
\label{sec-ffiso}
In the absence of a magnetic field and under the point-mass boundary  
condition, all equilibria are perfect spheres. 
The entire class of such isolated configurations bounded by a constant 
surface pressure was found by Ebert (1955) and Bonnor (1956).  The 
addition of a magnetic field satisfying our prescribed boundary 
conditions (eq.\ \ref{eq-abc}) does not change this situation; 
i.e., each sphere is still an exact, equilibrium state. The reason 
is that a uniform, parallel field at the background value exerts no 
internal forces.  Thus, isolated, spherical clouds should be 
magnetically force-free states.

As a first test of the $q-$method, we used it to construct the spherical
sequence. For $R_0 < R_{0,\mx}$, initial guesses of $\psi (r,z) = 0,~ 
A(r,z) = r/2$ led quickly to convergence.  For the higher$-\rho_c$ 
models beyond the radius peak, we again chose $A(r,z) = r/2$ and used 
the exact, nonmagnetic solutions with higher $\rho_c$ as initial guesses 
for $\psi$.  
Members of the resulting sequence all contained uniform fields. 
For example, we found that $B_c \equiv \vert {\bf B} (0,0) \vert$ 
never differed fractionally 
from unity (i.e., the background value) by more than $2 \times 10^{-2}$. 
We also checked that the density profile of each configuration matched 
that of its nonmagnetic, isothermal counterpart having the same central 
density $\rho_c$.  Thus, spherical clouds under the point-mass boundary 
condition are indeed members of the Bonnor-Ebert sequence.  

\subsubsection{Oblate Equilibria}
\label{sec-obeq}
More general, nonspherical, configurations with shapes given by equation 
(\ref{eq-shape}) require magnetic support.  To offset the weight of the
extra mass in the equatorial region, an oblate cloud has a field that 
rises inward, i.e., $\p B_z/\p r < 0$ for all $r$ and $z$. We used the
$q$-method to construct a number of states with axial ratio
$Z_0/R_0 = 1/2$. Our purpose was mainly to compare results with those  
of previous authors who employed the point-mass boundary condition
(Mouschovias 1976a,b; Tomisaka et al.\ 1988a,b).

Figure 3$a$ shows a representative equilibrium with $\alpha = 10$. 
As expected, the field lines have a modest inward bending toward the 
polar axis.  The horizontal arrows in the figure represent 
the drift velocities in this initial state, and show the 
instantaneous pattern of mass redistribution in the cloud.  All of 
these velocities point inward.  The peak drift speed $v_{d,\mx} = 
0.033$ occurs at the cloud equator (i.e., $r=R_0, z=0$), where 
the magnitude of the current also has a maximum.  

Figure 3$b$ displays the mass-to-flux
distribution $dm/d\Phi$ as a function of the magnetic flux $\Phi$. 
Because the density rises substantially toward the center, $dm/d\Phi$ 
peaks toward the axis ($\Phi=0$).  
We note again that previous authors specified this function ab initio. 
For example, Tomisaka et al.\ (1988a) used the $dm/d\Phi$ from a 
uniform-density sphere threaded by a constant magnetic field. 
This $dm/d\Phi$ also rises inward, but lacks the sharp central peak 
seen in our model.  The resulting cloud is oblate, but more ``boxy'' 
in appearance than ours. (See Fig.\ 2$d$ of Tomisaka et al.\ 1988a.)

In constructing the entire sequence of oblate equilibria, it is 
necessary to include states with higher central density than that 
at the radius peak $R_{0,\mx}$ (\S \ref{sec-eqseq}). 
To find these higher$-\rho_c$ states, we proceeded as follows. 
Taking a previously converged state with $R_0 < R_{0,\mx}$, we first 
increased $dm/d\Phi$ by a small, fixed amount at every $\Phi$.  This 
altered $dm/d\Phi$ was then used in the free-boundary method (\S 
\ref{sec-constr}) to converge 
a new state.  The resulting equilibrium, which was only roughly 
ellipsoidal, had the same total flux $\Phi_0$ as the original cloud, 
but a slightly larger mass and central density. Now using this state 
as an initial guess, its $dm/d\Phi$ was increased again, resulting in 
a cloud of still higher mass and central density.  The exercise 
was repeated several times, until $\rho_c$ significantly exceeded that 
of the last state converged via the $q-$method.  This high-density, roughly 
ellipsoidal model then served as an initial guess for the construction, 
now via the $q-$method, of a truly ellipsoidal state lying well past the 
radius peak.  In a typical case, the highest-density guesses constructed 
in this manner sufficed to converge equilibria that lie not only beyond 
$R_{0,\mx}$, but beyond $M_\mx$ as well. 

Figure 4 displays the masses of our oblate clouds as a function of their
central density, for $\alpha$-values of 1 and 10. 
Note that in these and all subsequent equilibrium sequences, the
total flux threading the cloud is {\it not} fixed a priori, but
varies continuously along the sequence. The dotted curve shows,
for comparison, the equivalent plot for the spherical, force-free states.
It is evident that the mass of any oblate configuration is significantly
greater than that of a force-free state with the same central density. For
example, the $\alpha = 10$ sequence has a peak mass of $M = 93.3$, attained 
at $\rho_c = 10.6$. This mass is 1.77 times the corresponding spherical 
value, which occurs at $\rho_c = 14.0$.  Note that while the central 
density corresponding to the peak mass is certainly smaller than in the 
spherical case, the broadness of the peak and finite sampling of the 
numerical calculations prevent an accurate determination of $\rho_c$ 
by the present method; our estimate may be in error by as much as
5--10 percent. 

The increased mass of oblate configurations has been noted by other 
authors (beginning with Mouschovias 1976b), and is readily understood. 
As Figure 3$a$ indicates, 
the actual bending of the magnetic field is relatively slight in all our 
models. Hence, force balance in the vertical direction involves mainly 
gravity and thermal pressure. Suppose now that we envision building up an 
oblate cloud by adding mass equatorially to an initially spherical 
configuration. Then $\rho_c$, which is largely determined by the weight 
of the central column, rises only slightly even for significant mass 
addition. By the same token, it is clear that configurations even more 
oblate than ours will have correspondingly higher masses at any central 
density.\fn{Any extra mechanical support beyond thermal pressure will
allow higher cloud masses. For the mass increase due to rotation, see
Figure 9 of Stahler (1983) and Figure 9 of Kiguchi et al.\ (1987).} 

One feature of the oblate models that may appear somewhat surprising 
is that their masses are insensitive to the background magnetic 
field strength, as parameterized by $\alpha$.  This is again a 
consequence of the relatively minor role played by the magnetic field 
in the vertical equilibrium.  When the vertical magnetic force is small, 
the polar radius is set almost entirely by $\rho_c$, as can be seen in 
the top panel of Figure 5.  This property, coupled with the fact that 
the cloud shapes are constrained to be identical, means that two clouds 
of the same $\rho_c$ must have nearly identical masses, whatever 
their values of $\alpha$.  

The above results imply that it is the shape of a cloud, not the 
strength of the magnetic field threading it, that determines the 
maximum equilibrium mass.  However, the degree of field line bending 
in a particular state {\it does} depend on the background field, as 
well as on the cloud shape.  We remarked
earlier that the extra equatorial mass in oblate clouds creates an
additional gravitational force in the radial direction. In response,
the magnetic field bends inward, until the associated tension helps 
offset this force.  The lower panel of Figure 5 quantifies the situation 
by plotting the central field value $B_c$ as a function of $\rho_c$.
For $\alpha = 1$, $B_c$ rises substantially, reflecting the increased 
field line bending in higher-mass clouds. This rise is much less for 
$\alpha = 10$, where the stronger field resists the equatorial gravity 
more effectively. 

\subsubsection{Prolate Equilibria}
\label{sec-pleq}

We next consider clouds that are prolate ellipsoids, with their long axes
parallel to the background magnetic field. 
Imagine creating such an object by shaving off equatorial mass from an 
initially spherical configuration.  By our previous reasoning, this change 
would only marginally affect the central density. Thus, a prolate cloud is 
less massive than a spherical one with the same $\rho_c$.  
In addition, the pressure gradient is higher in the radial direction than 
along the pole. Since the equatorial gravity is now insufficient to offset 
this extra force, the magnetic field 
bends outward, creating inward tension. The central field $B_c$ is reduced
from the background value, and $\partial B_z/\partial r$ is positive.
In this sense, isolated, prolate clouds are {\it magnetically confined}.

Using the $q$-method, we constructed sequences of prolate clouds with
axial ratio $Z_0/R_0 = 2$.  A representative state from the $\alpha = 1.5$
sequence is shown in Figure 6$a$.  The outward bowing of the field lines 
is evident, and is particularly strong near the center.
Here the field attains its minimum value, less than 10 percent of the 
background. The figure also shows the drift velocities associated with 
ambipolar diffusion.
These now point {\it outward}, reflecting the new field curvature. The
maximum drift speed, here equal to 0.3, occurs at the cloud equator.
We were not able to obtain equilibria with lower 
values of $\alpha$, most likely due to the very large gradient of 
$B$ at the origin as $B_c$ approaches zero. Note that, because $v_{d,\mx}$ 
is a significant fraction of the sound speed in these low-$\alpha$ states, 
our basic assumption of quasi-static equilibrium begins to fail. 
We shall revisit this issue in \S 4, when we attempt to follow explicitly 
the temporal evolution of prolate configurations.

Figure 6$b$ shows the mass-to-flux distribution in the cloud 
discussed above. Comparing with the analogous Figure 3$b$ for an oblate 
state, we see first that the maximum flux $\Phi_0$ is considerably smaller. 
This reduction is a consequence
of the prolate cloud's smaller cross section in the equatorial plane. We
also see that $dm/d\Phi$ climbs much more steeply toward the central axis, 
a result of both the increased column of gas near the pole and the reduced
flux in that region.

We display the cloud mass $M$ as a function of $\rho_c$ for the 
$\alpha = 1.5$ and $\alpha = 10$ sequences in Figure 7. As in the oblate
case, the curves are rather insensitive to $\alpha$. However, the deficit 
in equatorial mass now implies that $M$ is significantly less than the
force-free value at every $\rho_c$. Specifically, the mass is always below 
the Bonnor-Ebert value. The $\alpha = 10$ curve has the usual 
broad peak, now centered at $M = 37.8,~\rho_c = 14.9$.  This mass is 0.72 
times the corresponding maximum in the spherical sequence. For $\alpha = 
1.5$, however, the
central field becomes so small that we were unable to construct models for
$\rho_c$ beyond 15.5. Prolate clouds thus have a maximum mass 
set by the background field strength, if the latter is sufficiently low. 

The top panel in Figure 8 shows the polar radii of all our configurations, 
again as a function of $\rho_c$.  The insensitivity with respect to $\alpha$
is apparent, and demonstrates once more the minor role of the magnetic
field in vertical force balance.  The polar radii are now uniformly greater 
than their force-free counterparts.  Finally, the bottom panel of Figure 
8 shows in detail how the central field diminishes as the field lines 
bow outward. This curvature, and hence the drop in $B_c$, is relatively
small for $\alpha = 10$. On the other hand, $B_c$ rapidly falls to zero
in the $\alpha = 1.5$ case, where the curvature is much more pronounced.   

\subsection{Tidally Stressed Clouds}
\label{sec-tidal}
\subsubsection{Force-Free States}
\label{sec-fftid}
Under the periodic boundary condition, magnetically force-free 
configurations are no longer spherical.  Our stipulation that 
$\p \psi/\p z$ vanish at $z = \pm Z$ accounts for the gravitational 
influence of external masses located above and below the cloud of 
interest.  Their tidal force distorts the cloud vertically, i.e., in 
the direction of the uniform magnetic field.  The resulting equilibria 
are mildly prolate, although not perfectly ellipsoidal.  At fixed $Z$ 
and $R$ (the dimensions of the bounding cylindrical surface), there is 
a unique sequence parametrized again by $\rho_c$. The cloud axial ratio 
now varies along this sequence.

Figure 9 shows isodensity contours for three equilibria in a representative
sequence. Here, we have chosen $Z = 2.9$ and $R = 10$, following the 
discussion of \S \ref{sec-param}.  The clouds are nearly spherical at low 
$\rho_c$, as shown in the first panel. They become maximally distorted when 
the cloud's polar axis $Z_0$ peaks. The central panel of Figure 9 shows 
this pivotal configuration, at which $Z_0$ has reached 2.39, 
$Z_0/R_0 = 1.43$, and $\rho_c = 3.9$.  At higher central density, 
the polar axis shrinks and the cloud becomes more spherical again. 
The third panel is an example of such a high-density state.

If we view the cloud as one in a periodic chain, then the value of $Z$ 
represents half of the intercloud spacing. It is hardly surprising, then, 
that the character of our force-free sequences is sensitive to $Z$.
The choice $Z = 2.9$, as we have seen, results in a sequence for which 
the maximum $Z_0$ is close to $Z$ itself. For larger $Z$, the clouds are
less tidally distorted. For $Z \lae 2.8$ and sufficiently large $\rho_c$, 
the upper and lower boundaries of the cloud penetrate the cylindrical 
boundary, i.e., the individual 
clouds in the chain partially coalesce. Note that our results are much less 
sensitive to the value of $R$, at least beyond some minimum. For the $Z= 2.9$
case, for example, we found that raising $R$ to 20 had a negligible effect on
our models.
Interestingly, the masses of the clouds in the force-free, tidal sequence 
do not differ significantly from their counterparts of the same central 
density in the Bonnor-Ebert sequence; the agreement is better than 1 
percent throughout the range of $\rho_c$ examined.  

\subsubsection{Spherical Equilibria} 
\label{sec-spheq}
The properties of tidally stressed clouds of various shapes may be deduced 
from their relation to the force-free states. Thus, consider a true prolate
ellipsoid with a higher aspect ratio $Z_0/R_0$ than the mildly elongated, 
force-free state of the same $\rho_c$. The ellipsoidal configuration 
experiences less gravitational force toward its axis, and so has an 
internal magnetic field that bows outward. Its structure will be 
qualitatively similar to that of the prolate clouds studied in \S 
\ref{sec-pleq}.  By analogy, a highly oblate cloud has field lines 
that bow inward, again like its isolated counterpart. 

An interesting difference arises, however, when we study {\it spherical}
clouds. These are no longer force-free. In fact, their aspect ratio is 
{\it less} than that of the corresponding force-free states. Their field 
lines thus bend inward, and $\p B_z/\p r < 0$. During ambipolar diffusion,
the drift velocities point toward the symmetry axis, leading to states of 
higher central density. We are thus motivated to study these objects
further.  Historically, various authors have considered spherical, 
magnetized clouds as a convenient idealization (Mestel 1965; Strittmatter 
1966). However, these studies ignored the detailed field topology, as did 
the more recent evolutionary calculations of Safier, McKee, \& Stahler 
(1996) and Li (1998).

To construct the spherical sequence we again utilized the $q$-method, 
now with the tidal boundary condition, equation (\ref{eq-tidal}).
As always, the procedure was straightforward up to the radius peak. In 
converging states beyond this point, we used Bonnor-Ebert solutions of 
similar $\rho_c$ to supply an initial guess for $\psi$. We guessed $A$ 
by first constructing a slightly oblate cloud of nearly the same $\rho_c$, 
under the same tidal boundary condition.  Such a state has an inwardly 
bent magnetic field, similar to the tidally stressed sphere of interest. 
As we moved to higher-$\rho_c$ spherical states, the amount of field-line 
bending increased. Hence, we needed to use progressively more oblate 
clouds to supply the guess for $A$.

Figure 10$a$ shows the isodensity contours and field lines for a cloud 
from the $\alpha = 10$ sequence. The degree of field line bending and 
enhancement of $B_c$ are only slight for this relatively high 
$\alpha$-value. As expected, the drift
velocities point inward, with the largest velocity occurring at the 
equator. Both the field topology and pattern of drift speeds resemble those
for the isolated, oblate cloud of Figure 3, which has nearly the same central
density. The mass-to-flux distribution in the sphere, shown in Figure 10$b$,
is also qualitatively similar.

Figure 11 displays the mass of our spherical clouds as a function of 
central density, for $\alpha = 1$ and 10. The outer cylindrical boundary 
was again held fixed at $Z = 2.9,~R = 10$. The force-free sequence is 
represented by the dotted curve. For a given $\rho_c$, the magnetically 
supported states are evidently more massive than both force-free, 
tidally stressed clouds and Bonnor-Ebert spheres.  In the first case, 
the greater mass is a consequence of the objects' larger equatorial
extent, while the tidal boundary condition is responsible in the second.
More specifically, the vanishing potential gradient at $z = \pm Z$ 
effectively weakens self-gravity everywhere, so that a higher equilibrium 
mass is required at a given $\rho_c$ than for the point-mass boundary 
condition.
The maximum mass of 60.1 in the tidally stressed, spherical sequence 
exceeds the force-free and Bonnor-Ebert values by 13 percent. Note finally 
that the mass in our sequence peaks earlier ($\rho_c = 9.9$) than in 
either the force-free ($\rho_c = 15.6$) or Bonnor-Ebert ($\rho_c = 14.1$) 
case.

The top panel of Figure 12 shows the variation in radius along the 
$\alpha = 10$ sequence. Notice that the radii are intermediate between 
the force-free polar and equatorial values, shown by the dashed and dotted 
curve, respectively. The radius curve for the $\alpha = 1$ sequence is 
nearly identical to the $\alpha = 10$ result, and so is not displayed. 
Both curves lie slightly above the radii of Bonnor-Ebert spheres (compare 
Figure 5). In the lower panel of the figure, we display the variation in 
the central magnetic field value. The two sequences are now clearly 
distinguishable, with the $\alpha = 1$ curve having greater field bending 
and thus a higher $B_c$.

\subsection{Remarks on Dynamical Stability}
\label{sec-stab}
Figures 4, 7, and 11 all show the cloud mass reaching a maximum value
and then declining as $\rho_c$ increases. Such behavior is well known
in other contexts, where the mass extrema demarcate a stability
transition within the fundamental mode of oscillation, and thus the
onset of dynamical instability (Tassoul 1978). In addition to its
stellar applications, this ``static method'' for diagnosing instability
has also been used for rotating clouds (Stahler 1983; Kiguchi et al.\ 
1987). 

The situation is more subtle for magnetostatic configurations. The
physical basis of the static method is that two equal-mass equilibria
on either side of an extremum may be considered endstates in a normal
mode of zero frequency. For this interpretation to hold, however, the
two configurations must have identical distributions of all quantities
that are conserved during such an oscillation. For example, two rotating,
axymmetric states must have the same variation of specific angular
momentum, since the oscillation exerts no internal torques.

In the case of magnetized clouds, we may assume that flux freezing
holds during the dynamical oscillation (even one of zero frequency).
Thus, application of the static method requires that the mass-to-flux
distribution not change along the sequence. This is the case for
magnetostatic models constructed via the free-boundary method, i.e., 
with a specified functional form of $dm/d\Phi$ (\S \ref{sec-constr}). 
In our models, on the other hand, $dm/d\Phi$ varies continuously along 
the sequence, so the static method is not applicable.  While clouds of 
sufficiently high central density are certainly 
unstable, the actual transition must be displaced somewhat from the 
extremum. 

Another consideration comes into play for the prolate states. 
We have stressed that these equilibria, especially those of low
$\alpha$ and high $\rho_c$, are magnetically confined, with the field 
bowing outward in the central region (recall Fig.\ 6). In laboratory 
plasmas, such a field curvature tends to be dynamically unstable. If 
one interchanges two adjacent flux tubes, the energy of the system 
decreases. (For an elementary demonstration, see Section 8.3 of 
Nicholson 1983.) A normal mode analysis shows that the unstable 
perturbations have short wavelengths perpendicular to the magnetic 
field, and long wavelengths parallel to it (Freidberg 1987, p.\ 267). 
Self-gravity, neglected in the plasma context, should not materially 
affect the properties of such a mode.

\section{TIME EVOLUTION}
\label{sec-tevol}
The presence of a non-zero drift velocity in the states constructed 
in the previous section indicates that they will evolve in time via 
ambipolar diffusion.  Thus we next consider the quasi-static evolution 
of these clouds, as outlined in \S \ref{sec-quasi}.  We chose as 
initial configurations the oblate, prolate, and spherical clouds depicted 
in Figs.\ 3, 6, and 10, respectively.  The calculations in this 
stage employed the free-boundary method, rather than the $q-$method 
(\S \ref{sec-constr} and \ref{sec-quasi}).  As soon as the cloud was 
allowed to evolve, its shape immediately began to depart from a perfect 
ellipsoid. 
In those cases where the central concentration increased with time, 
the calculation terminated when the large density gradient near the 
cloud center made convergence difficult on our uniformly spaced grid.  
This typically occurred  when the maximum neutral velocity was 
0.2--0.5 times the sound speed.  Hence the termination roughly 
coincided with the breakdown of the quasi-static approximation 
and the beginning of dynamical collapse.

\subsection{Oblate Initial States} 
\label{sec-obevol}
We begin with the oblate cloud shown in Figure 3, which has an axial 
ratio of $Z_0/R_0 = 1/2$.  By use of the evolution equation 
(\ref{eq-dmdt2nd}), this state was evolved in time steps of magnitude 
$\Delta t = 0.05$, until the central density $\rho_c$ reached 115, 
at $t\,=\,t_{\rm fin}\,=\,11.23$.  Beyond this point, we could find 
no more equilibria.  We determined the endpoint accurately by reducing 
the time step to $\Delta t\,=\,0.01$ after $t\,=\,10$. The final time 
corresponds to 6.2~Myr, using the fiducial parameter values of Section 
\ref{sec-nondim}.  

Figure 13 shows the cloud at the final converged time, $t_{\rm fin}$.
Note the pronounced dip in the cloud boundary near the pole, a 
feature seen in previous oblate, magnetized models (see, e.g., Figure 
1 of Tomisaka et al.\ 1990).  This is a result of the increasing self-gravity 
in the central region, where gravity is unopposed by magnetic forces
along the field 
lines.  The neutral velocity increases toward the center, and  
the vectors become more radial. The maximum value of $|{\bf v}|$ is 
0.45, and occurs near the midplane at $(r,z) = (0.41,0.07)$.  

The increase of the cloud's central density and mass-to-flux ratio 
with time is shown in Figure 14.  As matter drifts into the central 
region, $(dm/d\Phi)_c$ rises steadily, increasing by 73 percent in 
the course of the evolution.  Simultaneously, 
the central field lines squeeze together, causing the value of $B_c$ to 
more than double by $t_{\rm fin}$.  Since the total flux decreases, the 
global distribution of $dm/d\Phi$ is becoming both narrower and more sharply 
peaked. The rapid increase of $\rho_c$, along with the large value of 
$|{\bf v}|$ near the center, both indicate that the quasi-static 
stage of evolution is ending. 

Finally, consider the evolution of the density profiles. At all times, the 
profiles
exhibit a turnover, leading to an envelope with nearly power-law density.
The radius of this turnover, i.e., of the inner core region, shrinks with 
time, while the profile in the envelope steepens. Both results are in accord 
with earlier studies (e.g., Mouschovias 1991).

\subsection{Prolate Initial States} 
\label{sec-plevol}
The unique feature of prolate structures is that they are magnetically
confined. That is, the field lines bow outward from the central axis,
creating the tension needed to oppose the larger pressure gradient in
the radial direction. Since the drift velocities also point outward,
we expect the evolution through ambipolar diffusion to differ
qualitatively from the oblate case.

We attempted to evolve the prolate initial state with $\alpha = 1.5$ 
shown in Figure 6$a$, but did not succeed.  As discussed in \S 
\ref{sec-pleq}, 
this state possesses a steep gradient of $B$ at the origin, which 
makes convergence difficult using the free-boundary method. 
Since this gradient is not as extreme for larger values of $\alpha$, 
we followed instead the evolution of a 2:1 prolate cloud with 
$\alpha = 10$.  Figure 15 shows the results.  The upper left panel 
displays the initial state. 
At early times, the outward neutral velocity causes the equatorial 
region to expand.  By the time $t = 5$ (2.8 Myr with our fiducial 
parameters), the central density has dropped to 11.2 and the maximum 
neutral velocity has decreased to less than half its initial value 
(see Fig.\ 15$b$).  During the same interval, the 
magnetic field has begun to straighten; its central value has increased 
slightly from 0.89 to 0.90. 

By the next time displayed in Figure 15$c$, $t = 20$ (11 Myr), the equator 
has moved out still further, while the pole has shrunk noticeably from 
its initial position. The cloud is clearly becoming {\it less} prolate, 
and now has an axial ratio of $Z_0/R_0 = 1.3$. The trends noted above in 
$\rho_c,~B_c$, and $v_\mx$ continue.  Because the field is straightening, 
ambipolar diffusion begins to slow down.  By $t = 50$ (28 Myr; 
Fig.\ 15$d$), where we stopped the calculation, the cloud is approaching 
a spherical shape ($Z_0/R_0 = 1.07$), albeit slowly, since the maximum 
neutral velocity is only 0.5 percent of the sound speed.  The central 
density has decreased to one-quarter of its starting value, so that 
the density profile is now much flatter than the initial one (see below). 
Over most of the cloud, the neutral velocities are oriented perpendicular 
to the field lines, and still point outward. Taken together, these 
developments indicate that the configuration will never reach the point 
of dynamical collapse. 

Instead, our calculation suggests that the endstate of a prolate,
magnetized cloud is a sphere threaded by a uniform magnetic field.  
Since a uniform field exerts no forces, the sphere would be 
indistinguishable from a non-magnetic, pressure-bounded configuration 
of the same mass as the initial prolate cloud. To test this idea, we 
constructed a sphere having the same central density ($\rho_c = 3.82$) 
as the prolate final state.  The corresponding mass of this non-magnetic 
state is (in our units) $M = 41.0$, which is only 5 percent larger than 
that of the state shown in Figure 15$d$.  Given 
that the axial ratio of this ``final'' prolate state is still 
7 percent greater than unity, this difference is understandable. 
Finally, we compare the mean of the equatorial and polar radii, 
$(R_p R_e^2)^{1/3}$, with the radius of an equilibrium sphere of 
the same $\rho_c$.  We find that the two quantities differ by
only 2 percent.

The evolution of the prolate cloud's central density and mass-to-flux 
ratio are shown in Figure 16.  The central density decreases smoothly from 
its initial value, asymptotically approaching the spherical, force-free 
value of $\rho_c = 3.8$ at large $t$.  The central mass-to-flux ratio also 
decreases steadily with time.  Once the final state is reached, 
$(dm/d\Phi)_c$ has decreased to one-third of its original value.  
Comparison of Figures 14 and 16 shows how the oblate and prolate clouds 
evolve in essentially opposite directions. Figure 17 displays the equatorial 
and polar density profiles of the initial, prolate state, and the same 
profiles in the final configuration. We also show the profile of a 
non-magnetic sphere with a central density of 3.8. In contrast to the 
oblate case, the core region {\it expands} with time, with the envelope 
gradient becoming significantly shallower by $t = 50$. 

\subsection{Spherical Initial States} 
\label{sec-spevol}
An isolated spherical cloud, i.e., one in which the gravitational 
potential obeys the point-mass boundary condition, 
is actually a member of the Bonnor-Ebert sequence (\S \ref{sec-ffiso}). 
Since the internal magnetic field is uniform and force-free, the drift
velocity should be identically zero, and the cloud should not evolve 
under ambipolar diffusion. We checked that this was the case. Thus, a
cloud with an initial $\rho_c$ of 14.0 increased its central density by 
only 2 percent over a period of 10 Myr. Its central magnetic field changed 
by 0.3 percent during this time, remaining very close to unity, while 
the maximum neutral velocity never exceeded 0.012.  Finally, 
the equatorial and polar radii never differed from the Bonnor-Ebert 
value by more than 2 percent.

The more interesting case is a spherical cloud constructed with the tidal
boundary condition. Here, the magnetic field bows inward, leading to 
drift velocities that point toward the central axis. Starting with the
configuration shown in Figure 10, we found that the central density
rose monotonically. We were able to follow the evolution until
t = 20.15, corresponding to 11.2 Myr in our fiducial units. By this
time, $\rho_c$ had increased from its initial value of 11.7 to 111.
Note that a comparable rise in $\rho_c$ for the isolated, oblate cloud took 
about half the time, as the drift velocities were roughly twice as large.

Imposition of the tidal boundary condition effectively weakens the
gravitational force in the vertical direction. Nevertheless, the climb in
central density causes the polar region to compress, just as in the 
isolated, oblate case. The pronounced polar dip may be seen in Figure 18,
which shows the density contours and magnetic field lines at t = 20.15. 
The pattern of drift velocities is also similar to the oblate case.  

Figure 19 displays the evolution of the cloud's central density and
mass-to-flux distribution. By the time the final state has been reached, 
$\rho_c$ has increased by almost an order of magnitude, $B_c$ has nearly
doubled, and the central mass-to-flux value has increased by 74 percent.
Once again, $\rho_c$ is undergoing a very sharp increase, a sign that
the dynamical phase of evolution is close at hand. The largest neutral
speed, which occurs near the cloud center at $(r,z) \simeq (0.33,0.07)$, 
has the value 0.16. 
The density profiles of the evolving configuration exhibit 
the same qualitative features as in the oblate cloud
evolution, i.e., a shrinking core region and a steepening envelope 
gradient.

\section{DISCUSSION}
\label{sec-discuss}
In this paper, we have taken the novel approach of specifying the cloud shape,
rather than the mass-to-flux distribution, in order to construct equilibrium
models. One of our significant findings is that these structures can be 
{\it either} oblate or prolate. Furthermore, we did not encounter, in our 
admittedly limited parameter survey, any restrictions on the aspect ratio in
either case. This result contrasts sharply with the traditional view, inherited
from Mouschovias (1976b), that magnetostatic equilibria are necessarily
flattened along the direction of the ambient field. We now see that this 
conclusion arose from the method of constructing models. 
Specifically, mass-to-flux distributions that plateau toward the cloud 
center yield flattened equilibria, while more centrally-peaked 
distributions result in elongated clouds.\fn{
Both our models and all previous ones assume that the cloud is axisymmetric
about the background field direction. Ward-Thompson et al.\ (2000) have
recently published the first observations of magnetic fields in starless
dense cores, using submillimeter polarimetry. In their three examples, the
projected field direction appears to be at an oblique angle with
respect to the cores'
long axes. This result, if confirmed by future studies, may indicate that
the assumption of axisymmetry requires modification (Basu 2000).}

We are not the first, however, to construct elongated equilibria. Tomisaka 
(1991) added a toroidal field component to the purely poloidal models of
Tomisaka et al.\ (1988a,b), and thereby obtained moderately prolate 
configurations.  In these models, the toroidal field exerts a radial pinch 
that aids gravity in squeezing gas toward the central axis. Fiege \& Pudritz 
(2000) have recently explored a large number of such models. Their clouds, 
which span a wide range of aspect ratios, are all of low mass and density 
contrast ($M \lae 0.8~M_{BE}, ~\rho_c \lae 11$, where $M_{BE}$ is the maximum 
mass of the force-free spherical sequence). 
Interestingly, a subset of their equilibria exhibit bowed-out field lines 
similar to those in our own prolate models (see, e.g., their Figures 2 and 7). 
In their case, however, the bowing is most pronounced near the cloud boundary, 
and only occurs when the toroidal field strength is comparable to, or 
exceeds, the poloidal one.  Because these twisted-field solutions 
have a non-zero toroidal component at infinity, the elongation of the
clouds depends in part on the magnetic field configuration throughout the 
larger, parent body. In our solutions, the prolate shape arises solely from 
the local mass-to-flux distribution. 

Our second major result is that the fate of a cloud under ambipolar diffusion
is determined both by its density contrast {\it and} by its shape. Figure 20
summarizes this finding for the case of isolated clouds. Here we have
reproduced, from Figures 4 and 7, the mass curves for oblate, prolate, and
spherical (force-free) configurations, where the first two have $\alpha = 10$.
The three open circles represent initial states whose evolution we followed 
numerically. 

Consider first the two oblate configurations. The initial state of higher 
density contrast has a mass larger than $M_{BE}$ ($M = 1.3~M_{BE};  
\rho_c = 2.95$). Since its field lines bend inward, the cloud's central 
density increases as it evolves, and the representative point moves 
toward the right.  Note that the contraction rate for this cloud is
much slower than for the example in \S \ref{sec-obevol}, which was at 
the mass peak of the oblate sequence.  
Initially, the equatorial region shrinks rapidly enough that the
aspect ratio $Z_0/R_0$ actually increases, from 0.50 at $t = 0$ to 0.74
at $t = 80$. Eventually, however, the polar dip becomes so pronounced that
$Z_0/R_0$ falls, reaching 0.55 at our last computed model ($t = 119$). At
this time, the central density is $\rho_c = 110$ and is climbing
rapidly. Thus, all oblate equilibria ultimately flatten before they 
collapse.

By contrast, the initially oblate cloud of lower density contrast 
shown in Figure 20 has a mass below $M_{BE}$ ($M = 0.86~M_{BE},~ 
\rho_c = 2.00$).  Although $\rho_c$ still increases with time, the 
configuration does not collapse, but reaches the force-free, 
spherical state 
marked by the filled circle.  By $t = 100$, the central density is 
3.58, within 25 percent of the predicted final value of 4.80, and the 
aspect ratio is $Z_0/R_0 = 0.94$, very close to 1.  Thus the cloud 
becomes less oblate and more spherical, while the field lines 
straighten out.

Finally, we turn to the prolate sequence, also shown in Figure 20.  
Since the mass curve now lies wholly below the spherical one, {\it no} 
initial state will lead to dynamical collapse. The example shown
by an open circle corresponds exactly to the state discussed previously 
in \S \ref{sec-plevol}.  There it was demonstrated that the central 
density falls as the configuration becomes more spherical, and that 
the cloud does not collapse.  
To summarize, all states move toward the force-free
curve in the diagram, but only those with $M < M_{BE}$ can actually reach 
the curve and so avoid collapse. An analogous situation holds for our 
spherical, tidally stressed equilibria. Referring again to Figure 11, we 
see that the initial state with $\rho_c = 11.7$ lies above the force-free 
curve, and so inevitably collapses.

Our third, and last, major result is that {\it none} of our initial equilibria,
whether isolated or tidally stressed, evolves in a manner consistent with
observed, star-forming clouds. As we reviewed in Section 1, the radio maps of
dense cores are best explained as an ensemble of elongated, three-dimensional
objects projected onto the plane of the sky. There is no appreciable difference
between the shapes of cores with or without internal stars (Jijina et al.\  
1999). Thus, a correct theoretical account should start with an elongated
configuration and have it remain elongated through the onset of dynamical
collapse.

The fault may lie in our neglect of external matter. In this study, we have
followed the conventional route of surrounding the cloud by a fictitious,
zero-density medium of finite pressure. Real dense cores, however, are embedded
within larger filamentary structures. This fact alone hints why the initial
states are elongated. During ambipolar diffusion, matter from the external
reservoir can flow down magnetic field lines, piling up until it is halted
by the internal pressure gradient. Thus, the polar flattening we obtained
may be alleviated or even avoided entirely. It will be interesting to see how
cloud evolution proceeds under these very different conditions. 
\\\\
\noi
The authors would like to thank Jason Fiege, Chris McKee, and Steve 
Shore for useful 
discussions.  This research is supported in part by a NASA grant to
the Center for Star Formation Studies.  The research of CC was
supported in part by an NSERC Postdoctoral Fellowship, while that of 
SS was funded by NSF Grant AST-9987266.

\begin{appendix}
\section{APPENDIX: NUMERICAL TECHNIQUES}
To implement the $q-$method, equations (\ref{eq-aeqnd}) and 
(\ref{eq-psieqnd}) were solved simultaneously on a uniform grid 
covering the upper right quadrant of the large cylinder; i.e., 
$0 \leq r \leq R,~ 0 \leq z \leq Z$.  We begin with an initial guess 
$A^{(0)}, ~\psi^{(0)}$, and a fixed cloud boundary $Z_\rcl (r)$. 
The function $q (\Phi)$ is then given by equation (\ref{eq-qmeth}), 
which we rewrite as 
\beq
q^{(0)}(\Phi) = P_0~{\rm exp}[-\psi^{(0)}_0/a^2],
\label{eq-qmethnum}
\eeq
with the superscript indicating the zeroth iteration. Recall that 
the subscript zero on $\psi$ and $P$ denotes their boundary values.  
As mentioned in Section 2, the specification of $q(\Phi)$ is
sufficient to calculate the right-hand side source terms of equations
(\ref{eq-aeqnd}) and (\ref{eq-psieqnd}).  At each
subsequent iteration, solutions for $A$ and $\psi$ subject to the 
boundary conditions (\ref{eq-abc}) and (\ref{eq-ptbc}) (point-mass) 
or (\ref{eq-tidal}) (tidal) were obtained 
to within an accuracy of $10^{-5}$.  These were used as provisional 
iterates $A_*^{(1)}, ~\psi_*^{(1)}$ in generating subsequent guesses, 
as explained below. At each step, the updated $\psi$ was then used 
in equation (\ref{eq-qmethnum}) to generate a new $q$. 

The number of grid  
points in the radial and vertical directions, $N_R$ and $N_Z$, for 
the different cloud shapes adopted in Section \ref{sec-init} were: 
\begin{center}
{\it Oblate:} $N_R = 81,~N_Z = 41$ \\
{\it Prolate:} $N_R = 41,~N_Z = 81$ \\
{\it Spherical:} $N_R = 61,~N_Z = 41.$
\end{center}
The number of field lines, $N_B$, was chosen equal to 
$N_R$.  To ensure accurate $\Phi-$derivatives near the cloud  
center, the flux value of the $i^{\rm th}$ field line was 
chosen as $\Phi_i = r_i^2/2,~ i = 1, 2, \ldots, N_B$.  In 
practice, it was necessary to switch back and forth between 
two interwoven meshes: one fixed spatially $(r,z)$, and another, 
defined by the current field line positions $(\Phi, z)$ 
(Mouschovias 1976$a$). 

The sequence of iterates was chosen according to the scheme:
\beqan
A^{(n+1)} = \theta_A~A^{(n)} + (1 - \theta_A)~A^{(n+1)}_* \\
{\rm and}~~~~~
\psi^{(n+1)} = \theta_{\psi}~\psi^{(n)} + (1 - \theta_{\psi})~
\psi^{(n+1)}_*,
\eeqan
where $\theta_A$ and $\theta_{\psi}$ are constant relaxation
parameters, taken to be between 0.5 and 1 in most of our calculations. 
A solution was deemed acceptable if the conditions
\beq
\left|\frac{A^{(n+1)}_* - A^{(n)}}{A^{(n+1)}_*}\right| < \eps 
~~~~~~~~{\rm and}~~~~~~~~
\left|\frac{\psi^{(n+1)}_* - \psi^{(n)}}{\psi^{(n+1)}_*}\right| < \eps 
\label{eq-converge}
\eeq
were satisfied simultaneously, with $\eps = 5 \times 10^{-3}$.  

We typically found solutions within 50 iterations or less, using 
an initial guess $\psi^{(0)} (r,z) = 0, ~A^{(0)} (r,z) = r/2$. 
We checked that the solutions were relatively insensitive to: (i) 
increasing the 
number of grid points in either direction; (ii) increasing 
the values of $R$ and $Z$; and (iii) allowing the program to continue 
for up to twice the number of iterations required to satisfy 
conditions (\ref{eq-converge}).  None of these resulted in 
a change of more than 5 percent in $\psi$ or $A$ at any grid point, 
or of more than 8 percent in the central density. 

The numerical implementation of the free-boundary method was 
discussed by Mouschovias (1976$a$).  In applying the method, 
we imposed an additional convergence criterion on the location 
of the cloud boundary:
\beq
\left|Z_{\rcl~*}^{(n+1)} - Z_\rcl^{(n)}\right| < \eps_z, 
\label{eq-zconv}
\eeq
where $\eps_z = \Delta z/2$.  Thus, at each timestep, convergence 
was achieved when conditions (\ref{eq-converge}) and (\ref{eq-zconv}) 
were simultaneously satisfied. 

In order to evolve the cloud over the hundreds of timesteps typical 
of our calculations in Section \ref{sec-tevol}, it was essential that 
the mass-to-flux distribution given by equation 
(\ref{eq-dmdphi2}) remain smooth throughout the evolution.  
To ensure this was the case, 
we approximated the function $q (\Phi)$ at each timestep by a 
least-squares polynomial.  Thus, $dq/d\Phi$ was also a 
polynomial, lending an additional degree of smoothness to 
equations (\ref{eq-dmdt2}) and (\ref{eq-dmdphi2}).   

\end{appendix}

{}

\bfig
\plotone{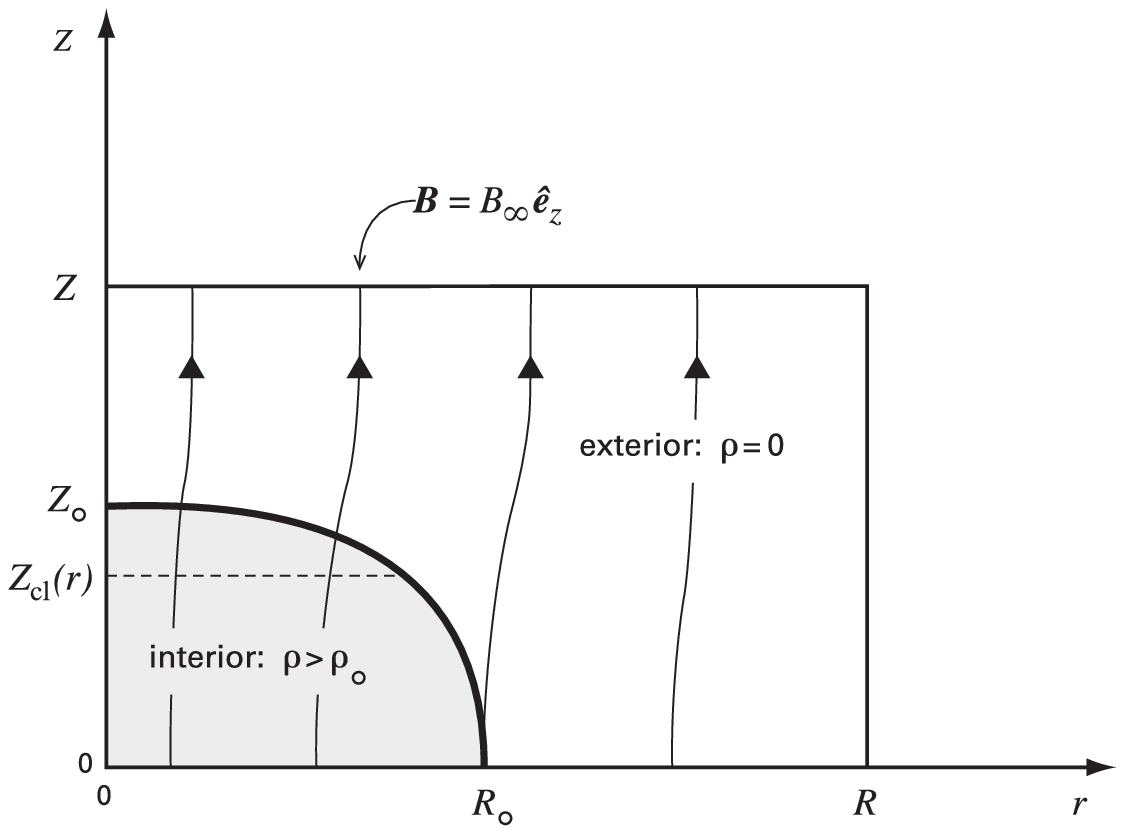}
\caption{
The computational volume is the upper right 
quadrant of a right circular cylinder of radius $R$, half-height 
$Z$.  The cloud is an axisymmetric structure with boundary $Z_{\rm 
cl} (r)$, centered on the origin, and embedded in a region of zero  
density but finite pressure.  Magnetic field lines in this 
figure are indicated by solid lines.} 
\efig

\bfig
\plotone{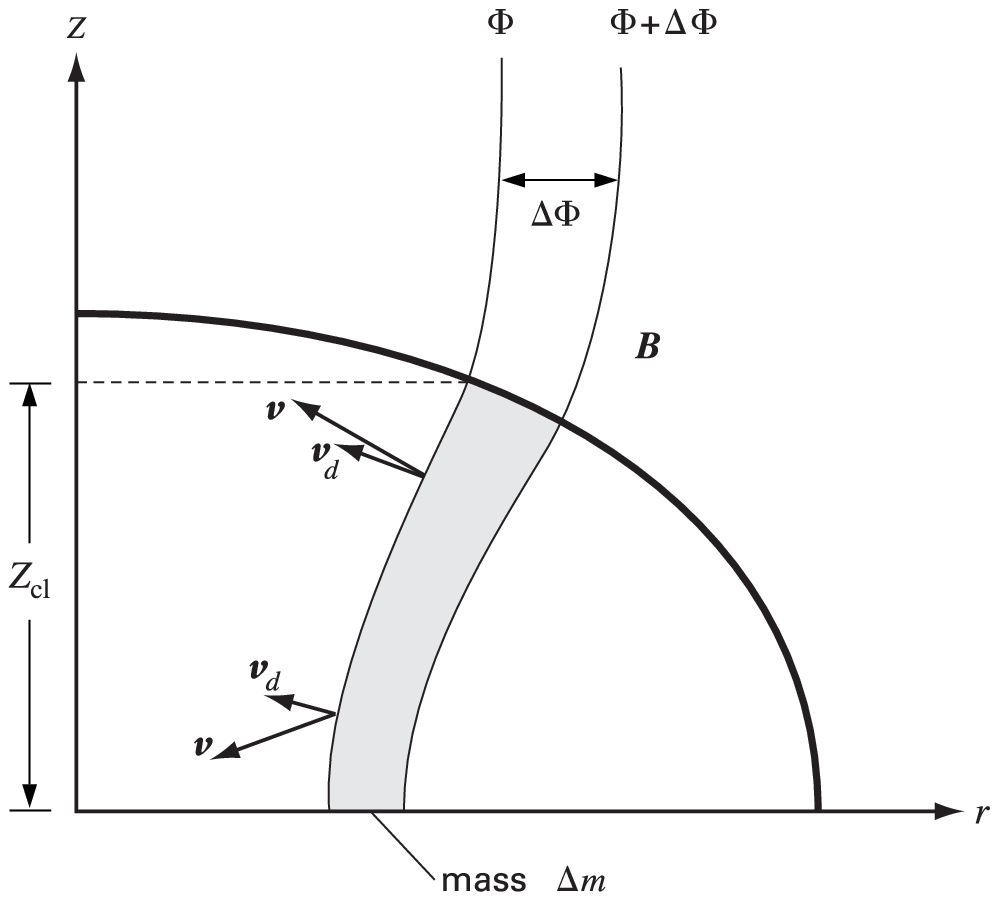}
\caption{The physics of ambipolar diffusion. A portion of the cloud, 
of mass $\Delta m$, is enclosed between two flux tubes, as shown. Over  
time, the neutral particles drift inward relative to the field with a
velocity ${\bf v}_d$. This drift velocity differs from ${\bf v}$, that 
of the neutrals in an inertial reference frame.}
\efig

\bfig
\plottwo{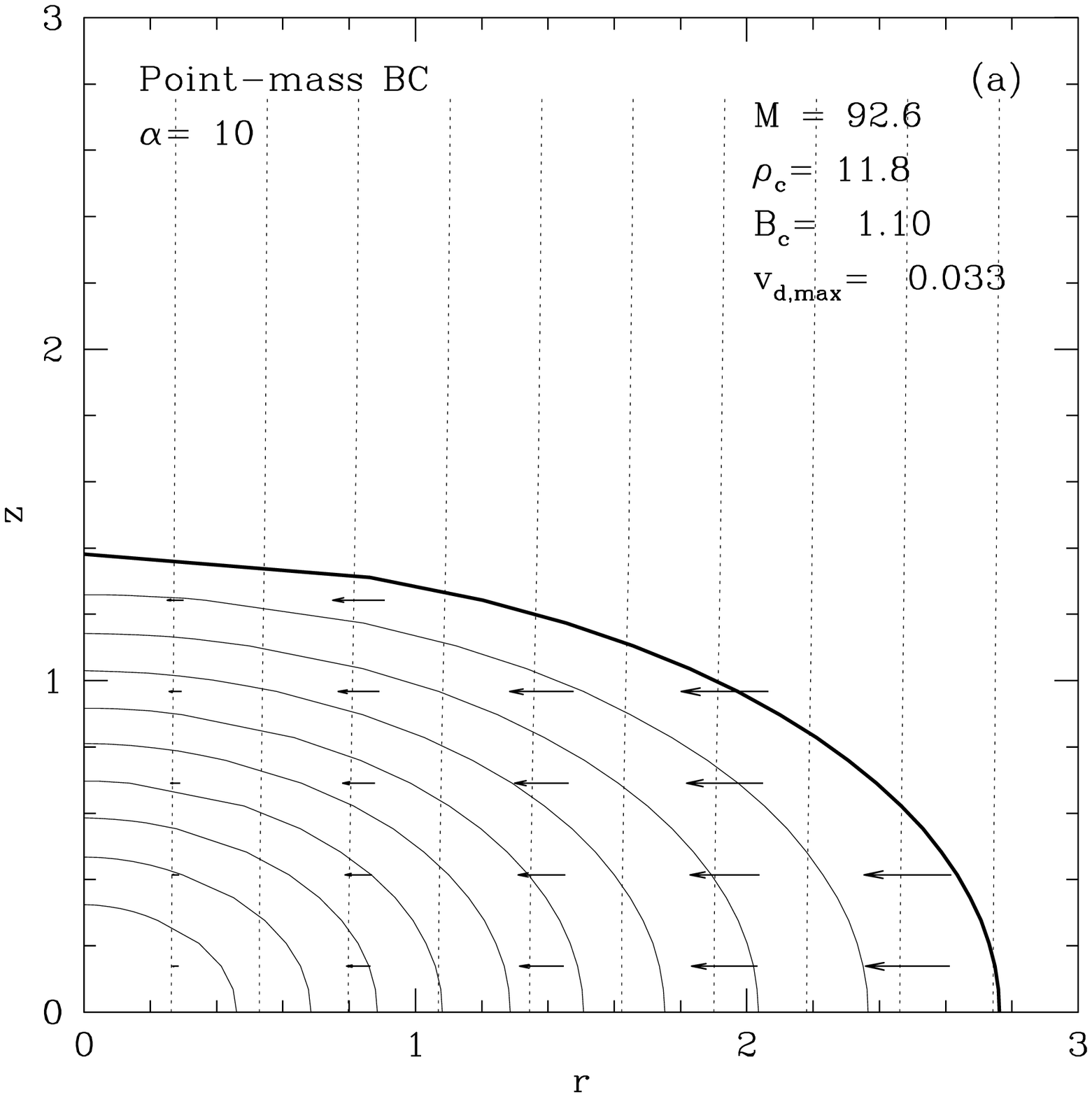}{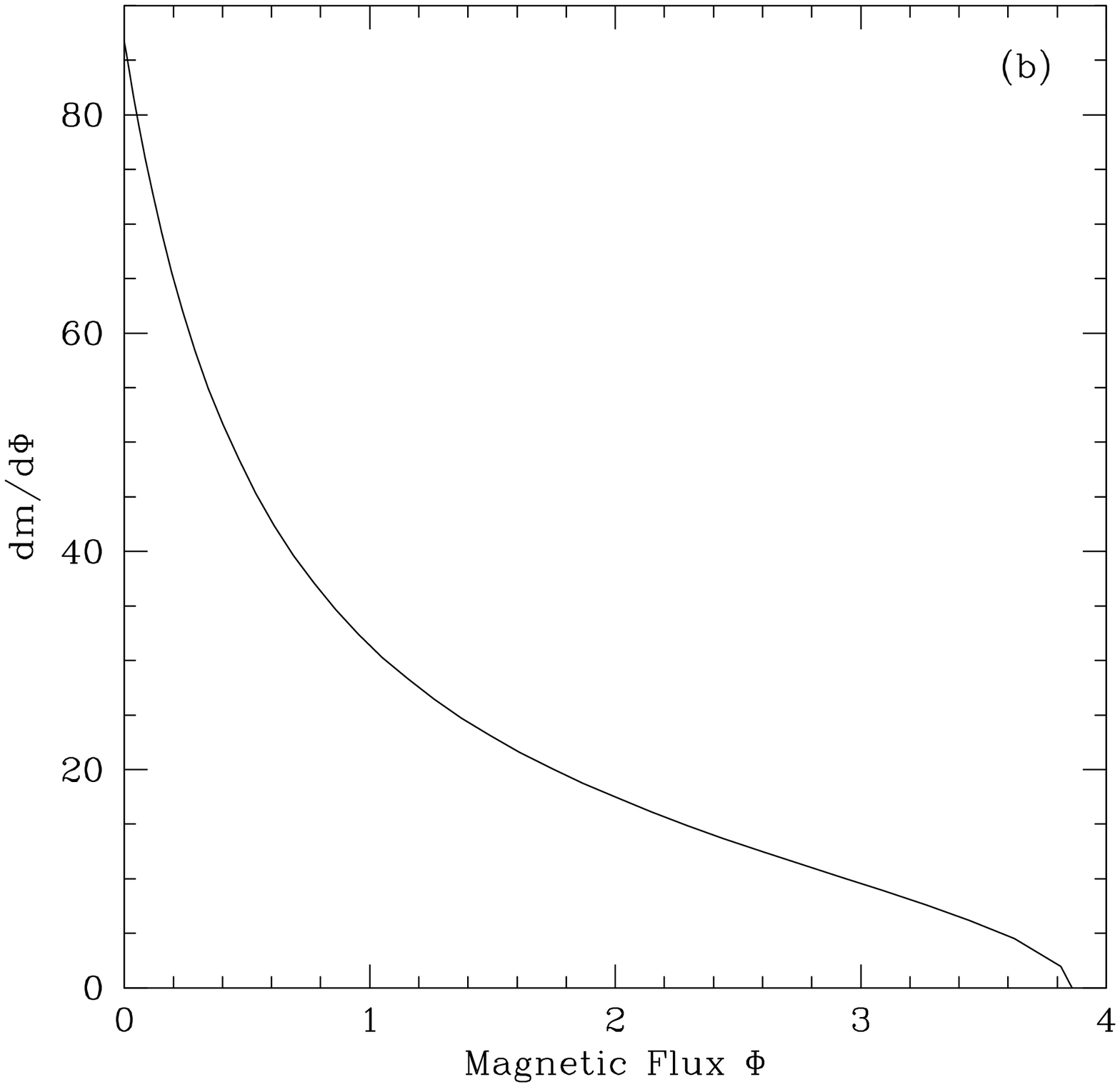}
\caption{(a) An oblate initial state with $\alpha = 10$.  Solid curves 
are isodensity contours and dotted curves are magnetic field lines. 
The heavy solid curve is the cloud boundary, which has an axial ratio 
$Z_0/R_0$ of 1/2.  Arrows indicate drift velocities, ${\bf v}_d = 
{\bf v} - {\bf v}_i$.  
The dimensions of the computational volume are $R = 5.52,~Z = 2.76$. 
(b) Mass-to-flux distribution as a function of magnetic flux $\Phi$ 
for the same cloud.  
\label{fig-obinit}} 
\efig

\bfig
\plotone{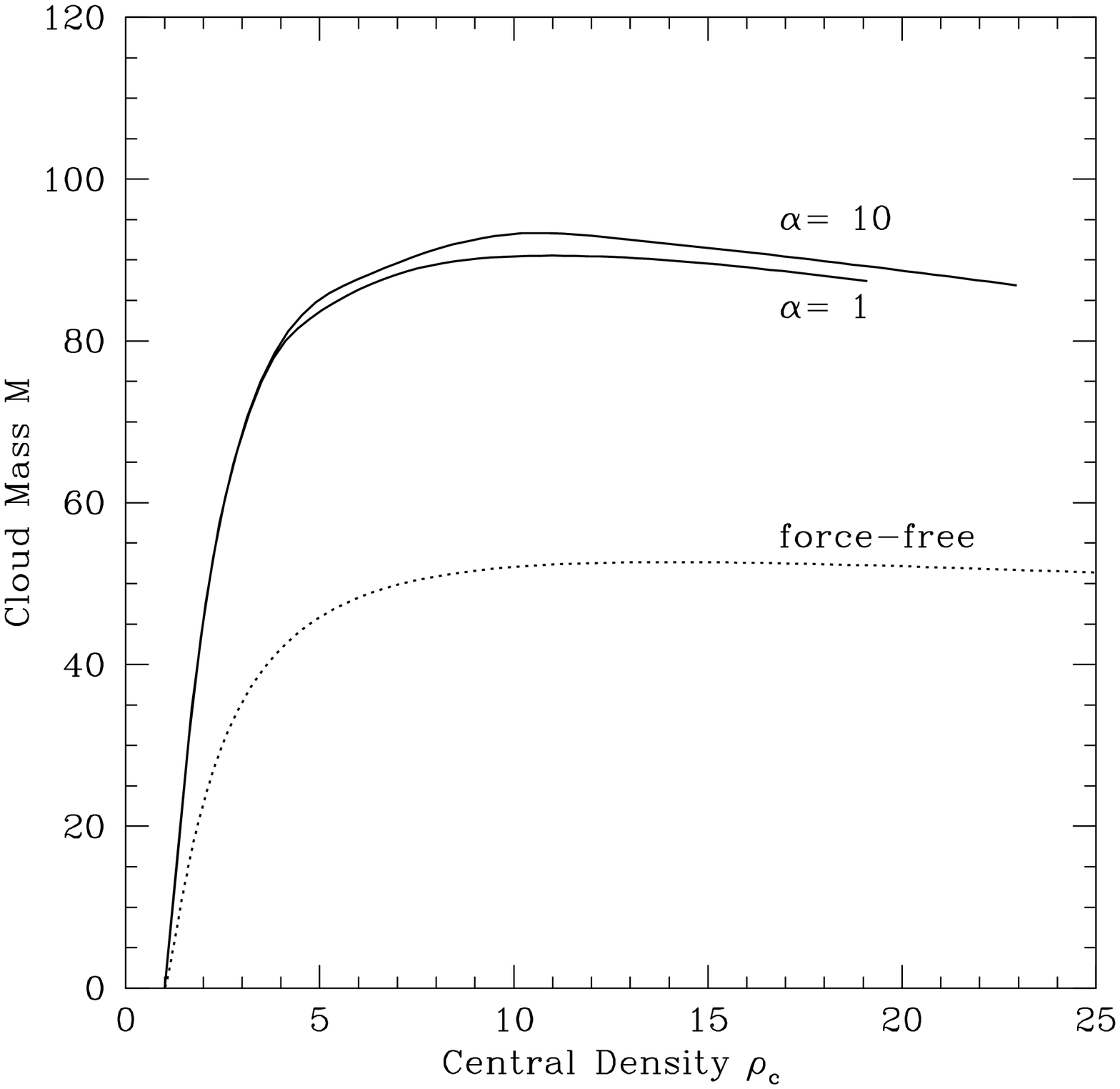}
\caption{
Mass as a function of density contrast (solid curves) for isolated, 
2:1 oblate equilibria.  Results for both values of $\alpha$ 
are displayed.  The corresponding force-free sequence of 
spherical (Bonnor-Ebert) equilibria is also shown (dotted curve).}
\efig

\bfig
\plotone{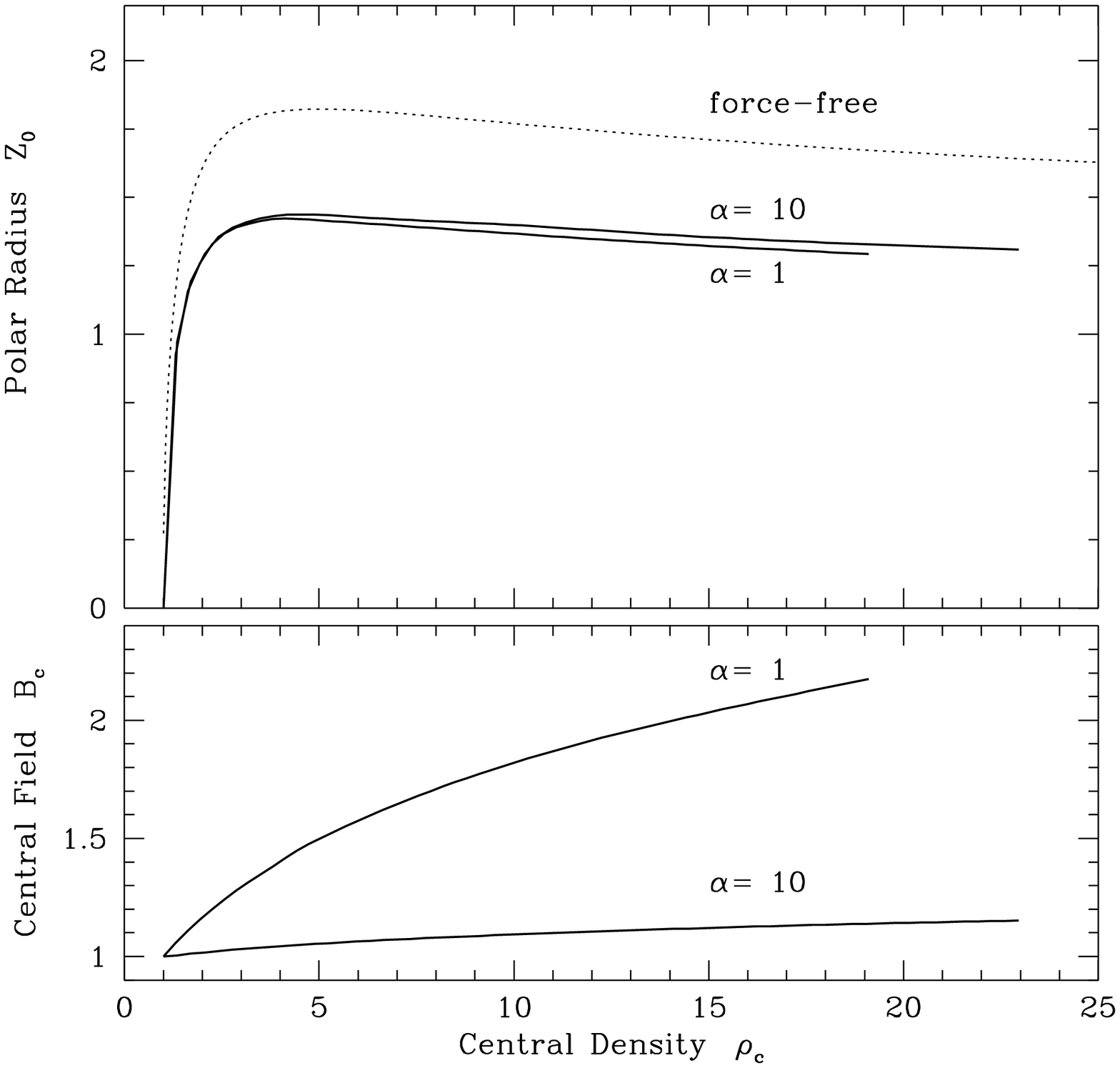}
\caption{{\it (top)} Polar radius versus density contrast 
for the oblate 2:1 sequence (solid curves).  The corresponding 
force-free sequence is also shown (dotted curve). 
{\it (bottom)} Central magnetic field versus density contrast, 
along the same sequences.}
\efig

\bfig
\plottwo{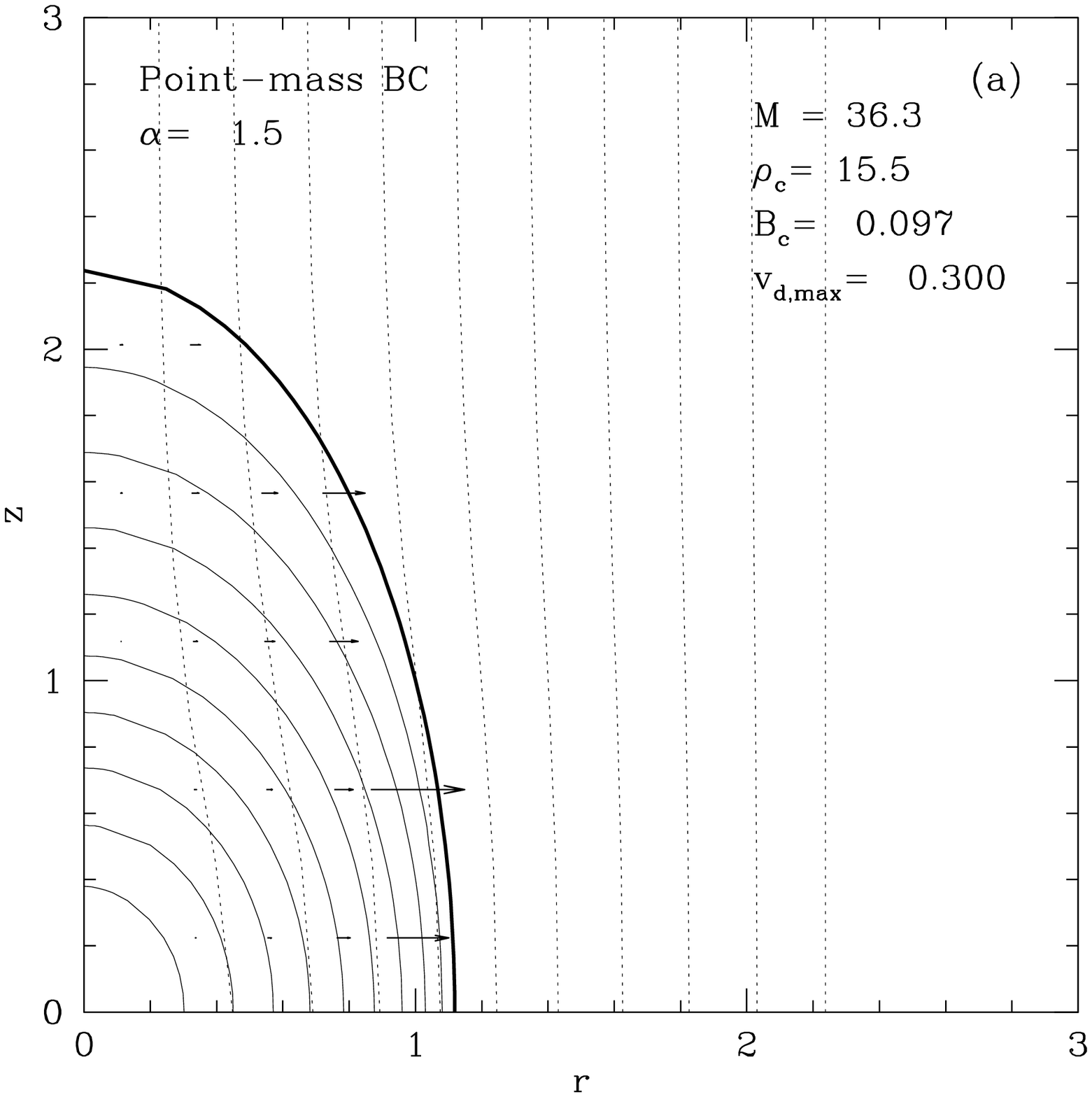}{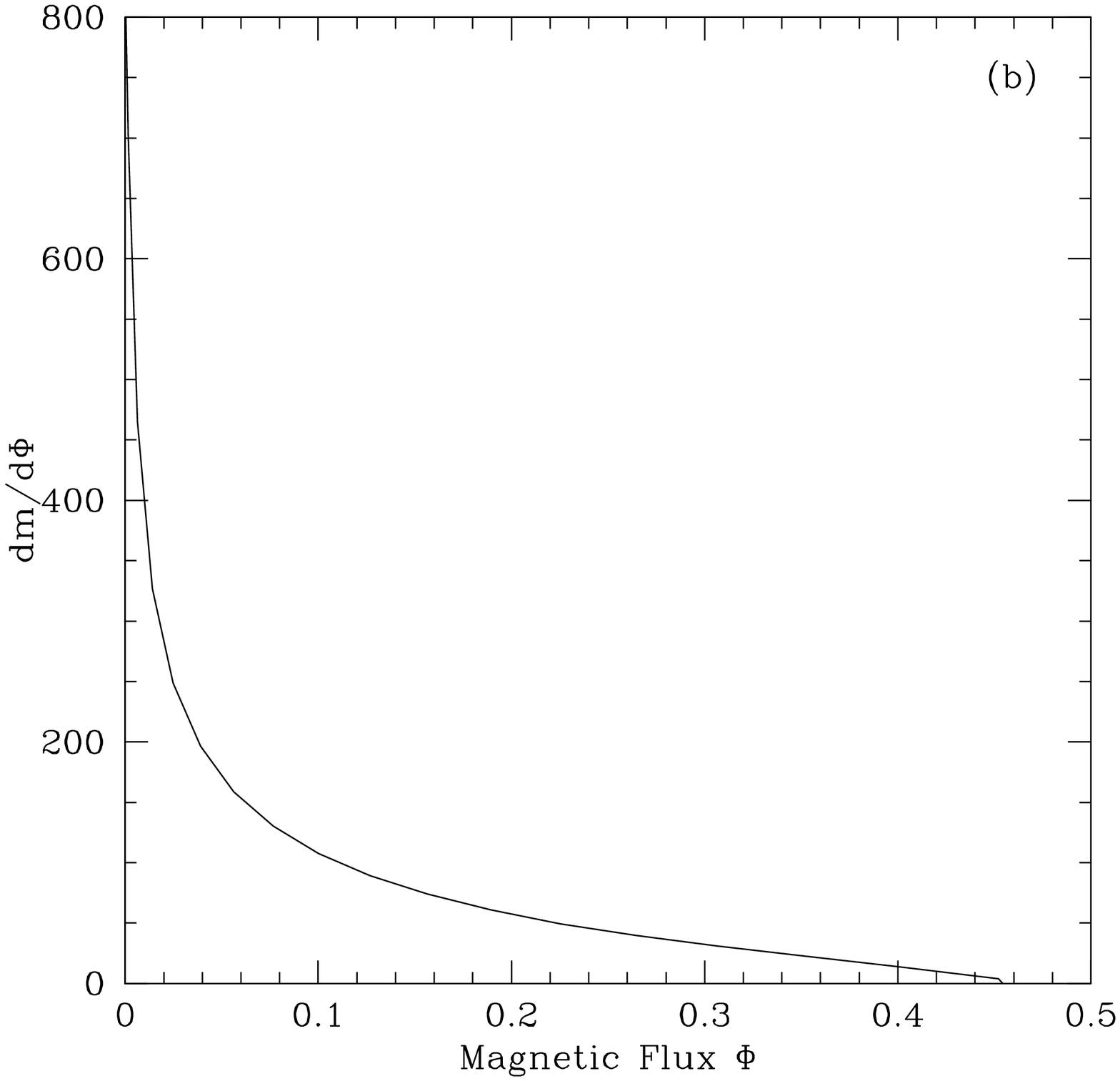}
\caption{
(a) An isolated, 2:1 prolate equilibrium cloud with $\alpha = 1.5$.  The 
magnetic field strength at the origin, $B_c$, is about 10 percent of its 
asymptotic value.  Here, the drift speeds point {\it outward} at all points
in the cloud.  The maximum drift speed, $v_{d,\mx}$, is 30 percent of the 
sound speed. The dimensions of the computational volume are $R = 2.24,
~Z = 4.47$. 
(b) The mass-to-flux function of the cloud in (a).} 
\efig

\bfig
\plotone{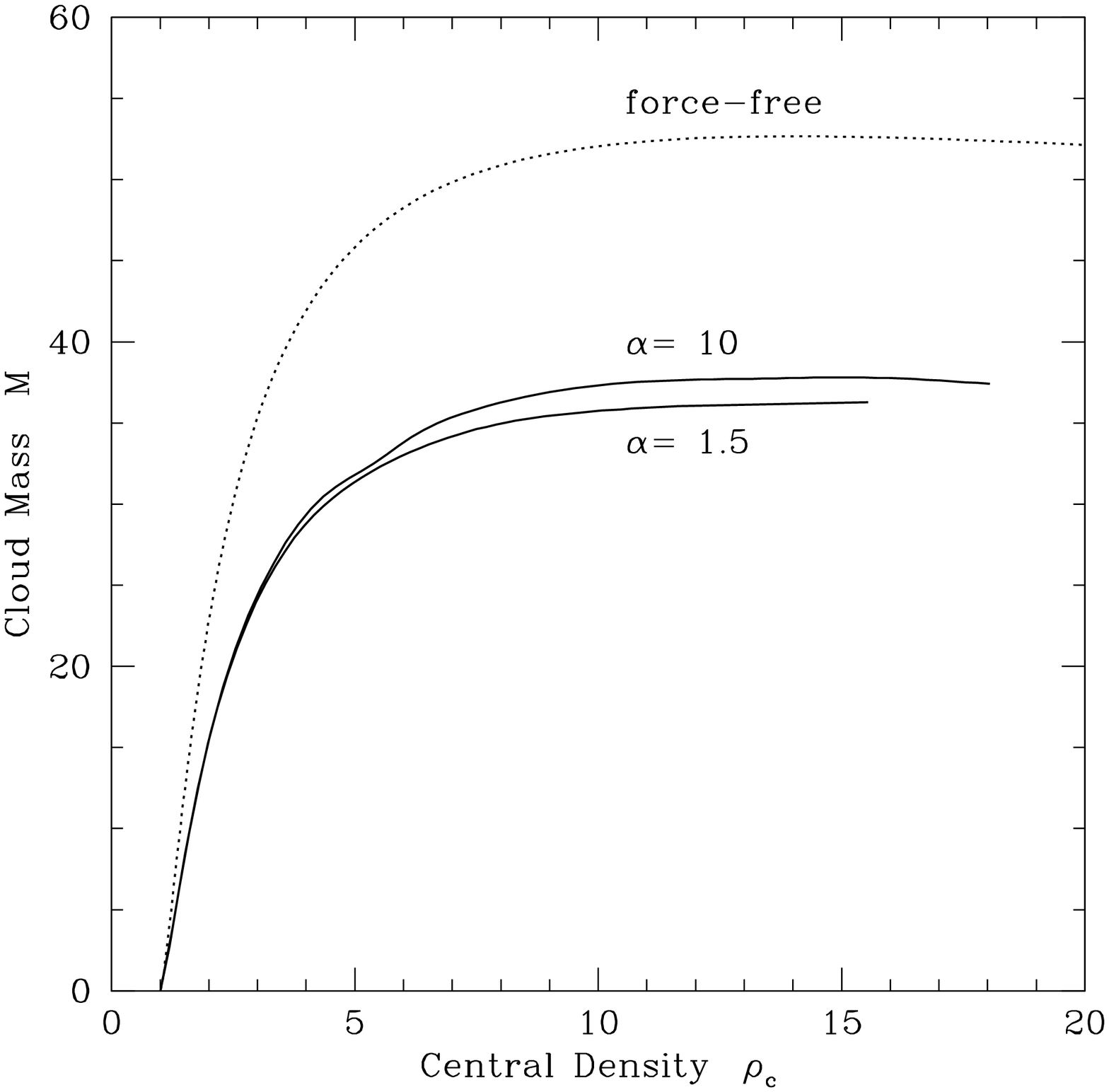}
\caption{
Same as Figure 4, but for the sequence of 2:1 prolate equilibria 
(solid curve).  The dotted curve shows the sequence of force-free 
(spherical) equilibria.}  
\efig

\bfig
\plotone{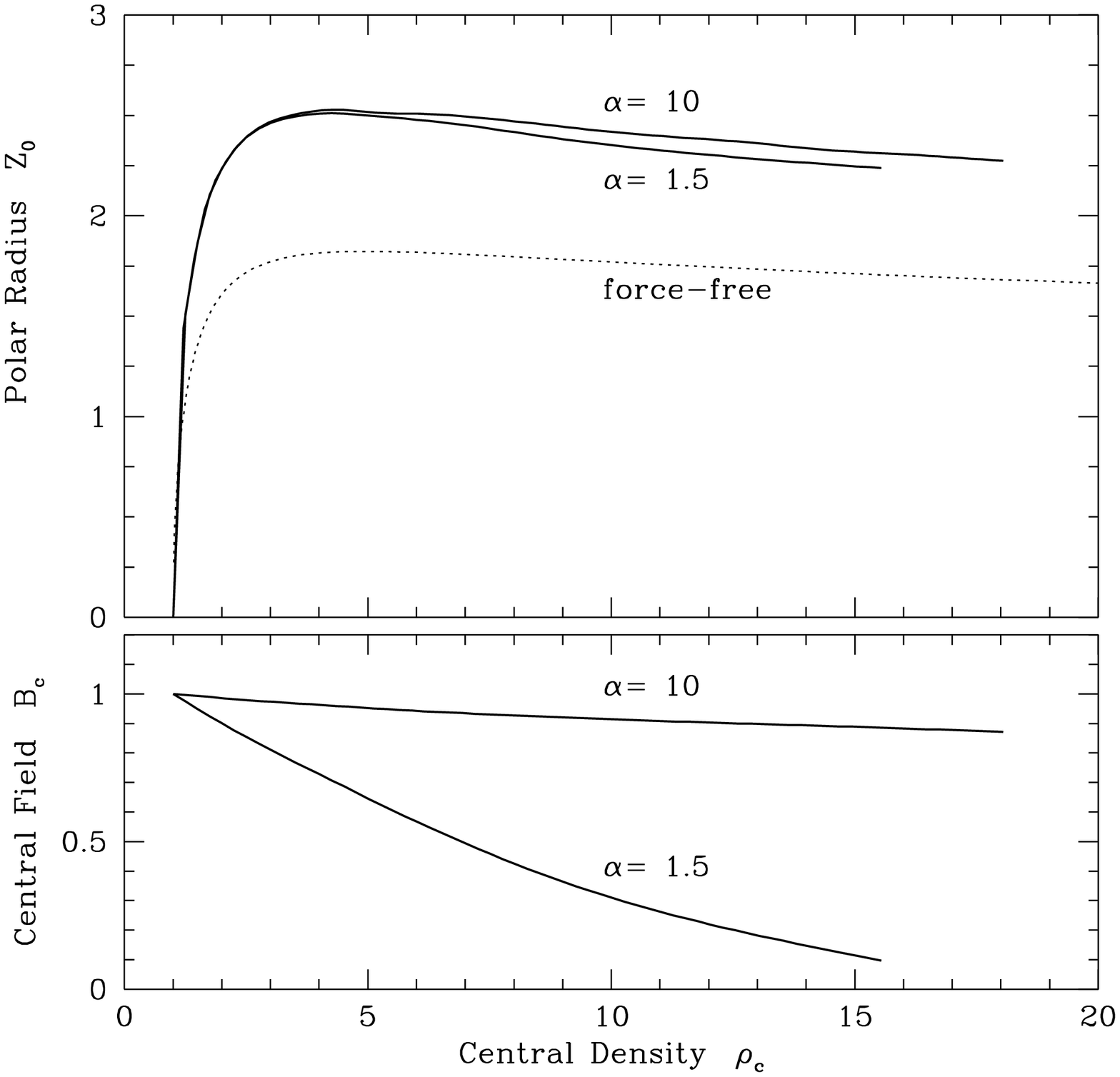}
\caption{{\it (top)} Polar radius versus density contrast 
for the prolate 2:1 sequence (solid curves).  The corresponding 
force-free sequence is also shown (dotted curve). 
{\it (bottom)} Central magnetic field versus density contrast, 
along the same sequences.}
\efig

\bfig
\plottwo{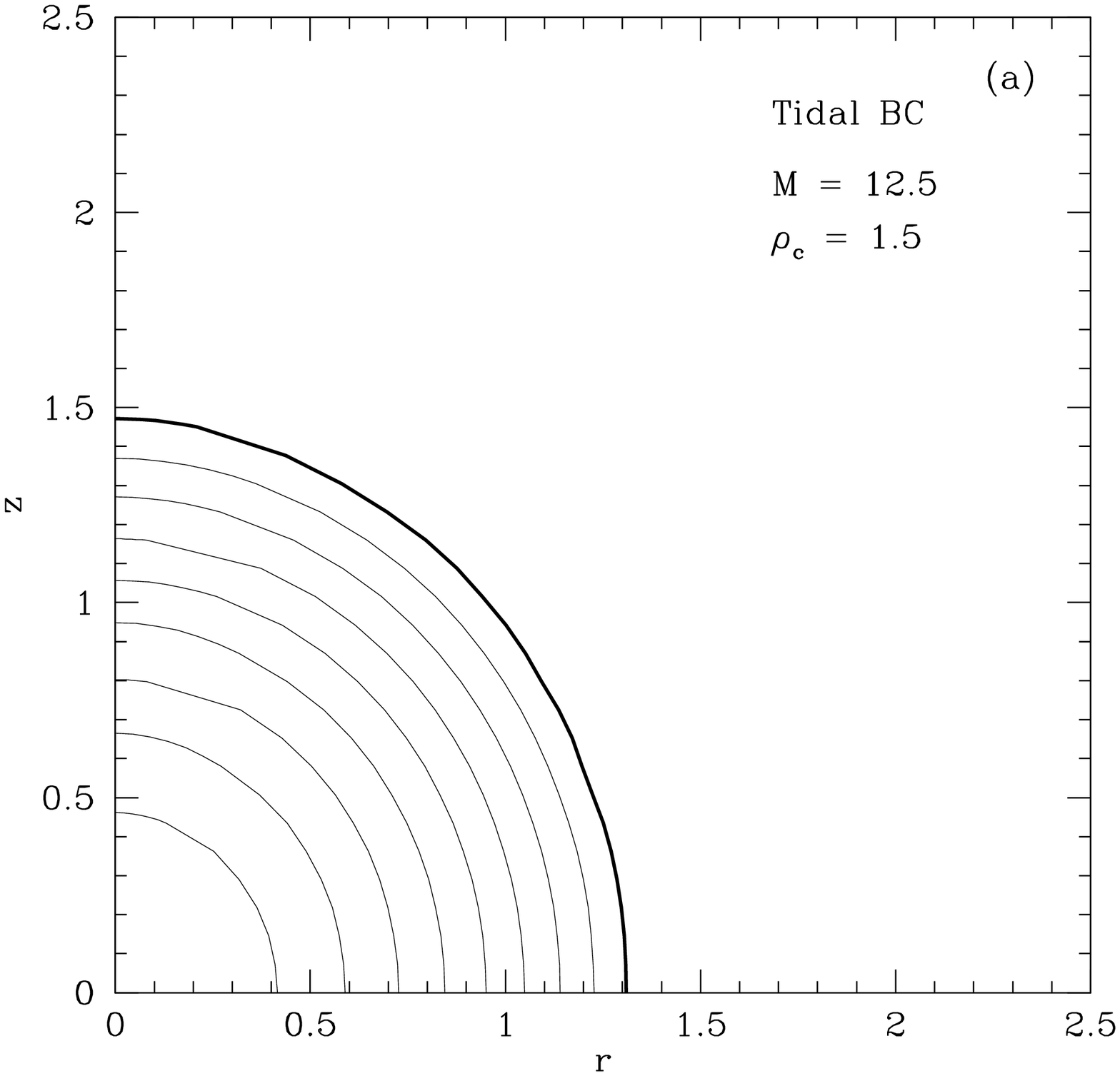}{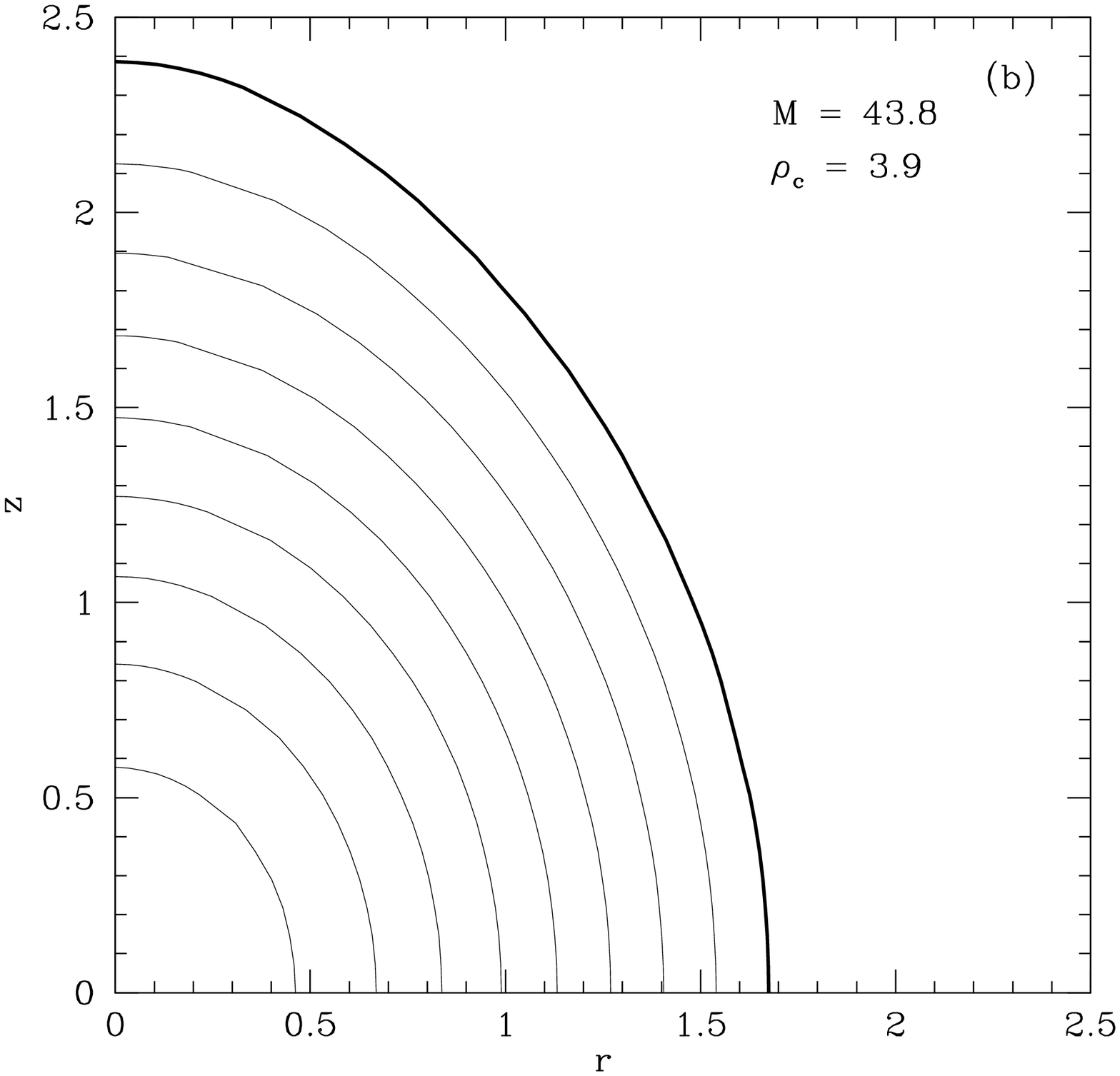}
\plotone{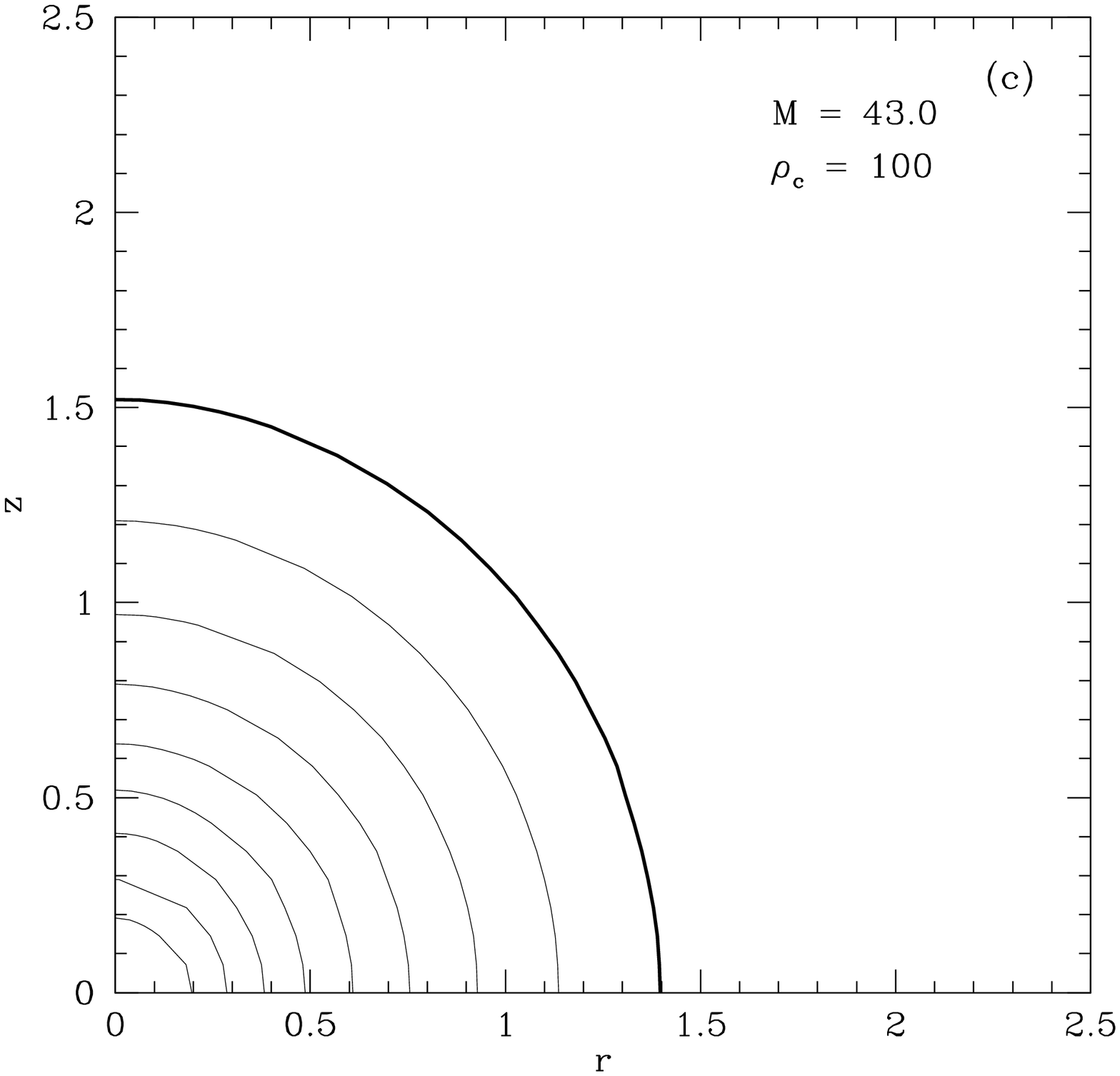}
\caption{
Force-free equilibria under the tidal boundary condition.  The 
dimensions of the computational volume are $R = 10.0,~Z = 2.9$. 
(a) A low-density cloud with $\rho_c = 1.5$. 
(b) Maximally distorted cloud with $\rho_c = 3.9$.
(c) A high-density cloud with $\rho_c = 100$.}
\efig

\bfig
\plottwo{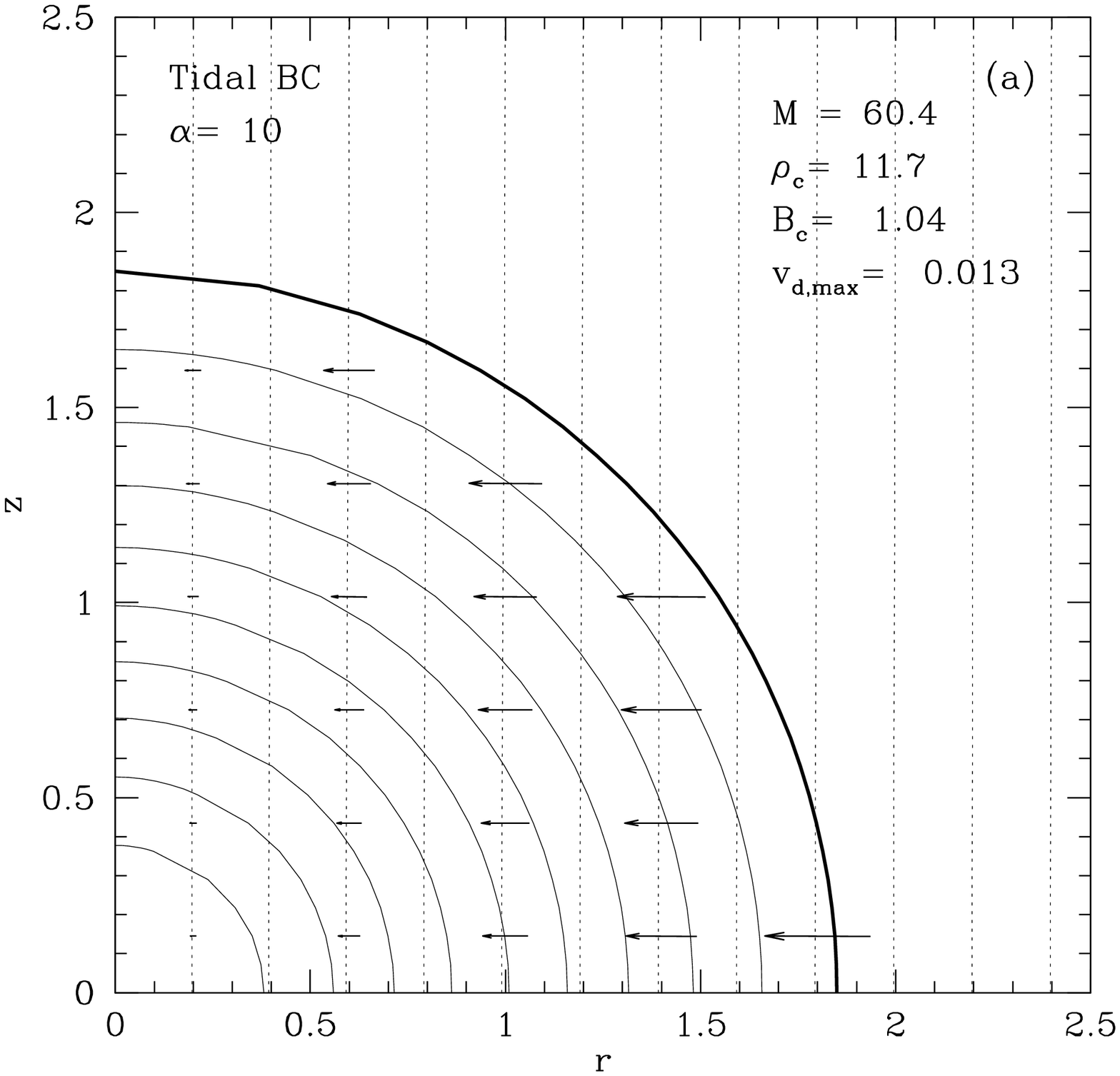}{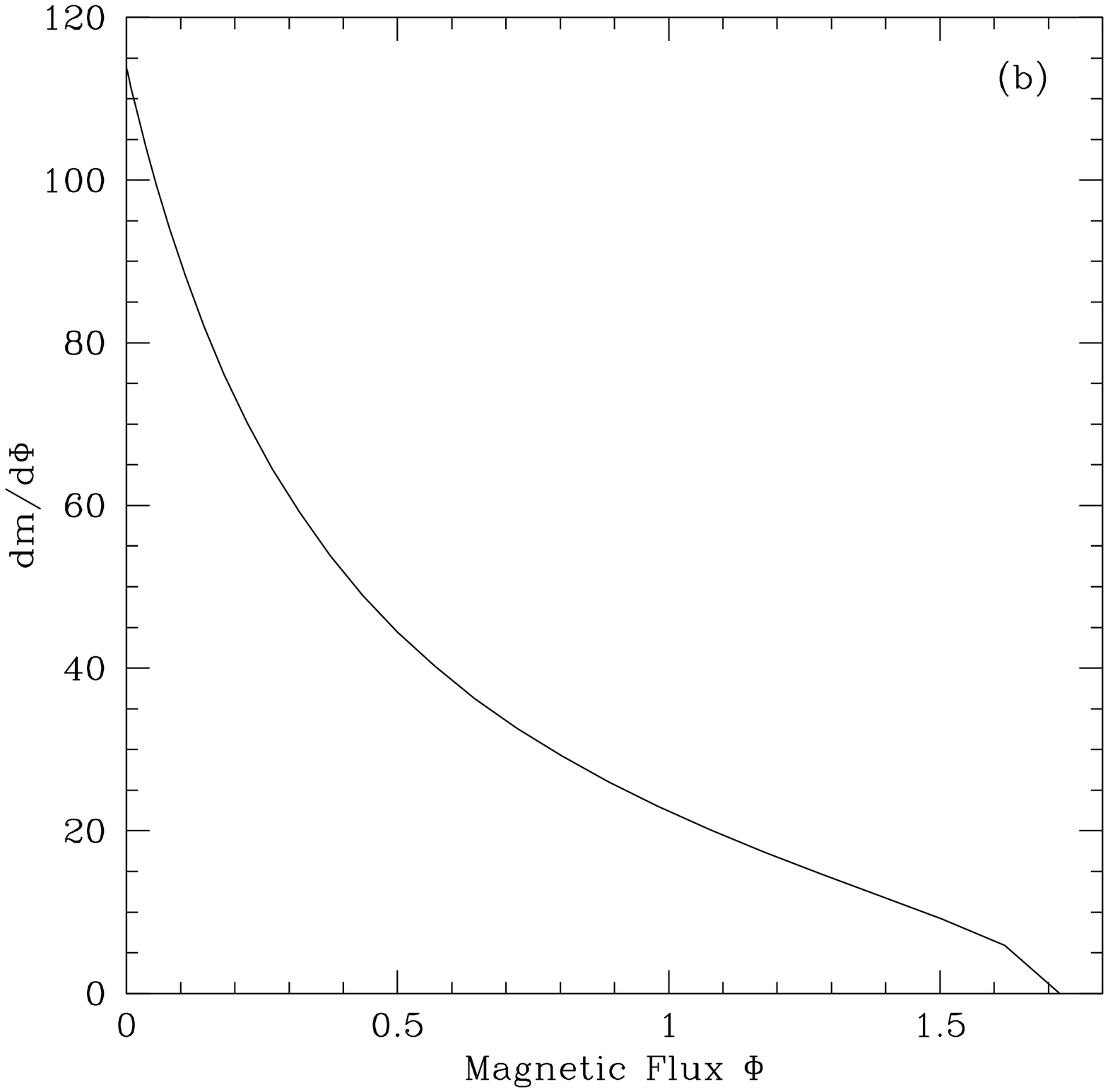}
\caption{
(a) A tidally stressed, spherical equilibrium cloud with $\alpha = 10$.  
The dimensions of the computational volume are $R = 10.0,~Z = 2.9$. 
(b) The mass-to-flux function of the cloud in (a).} 
\efig

\bfig
\plotone{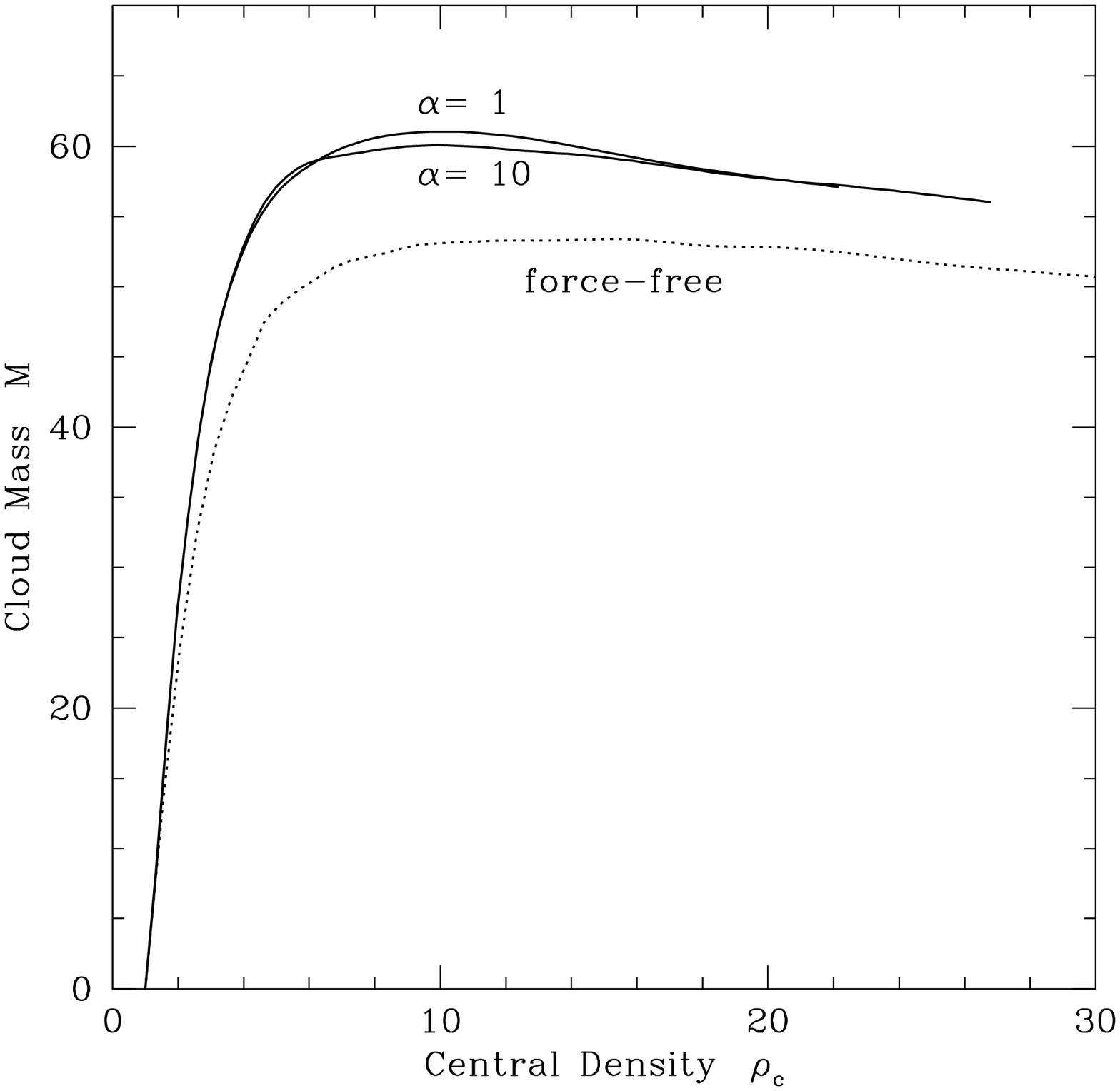}
\caption{
Same as Figure 4, but for the sequence of spherical equilibria under 
the tidal boundary condition (solid curve).  The dashed curve shows the 
sequence of force-free (prolate) equilibria.} 
\efig

\bfig
\plotone{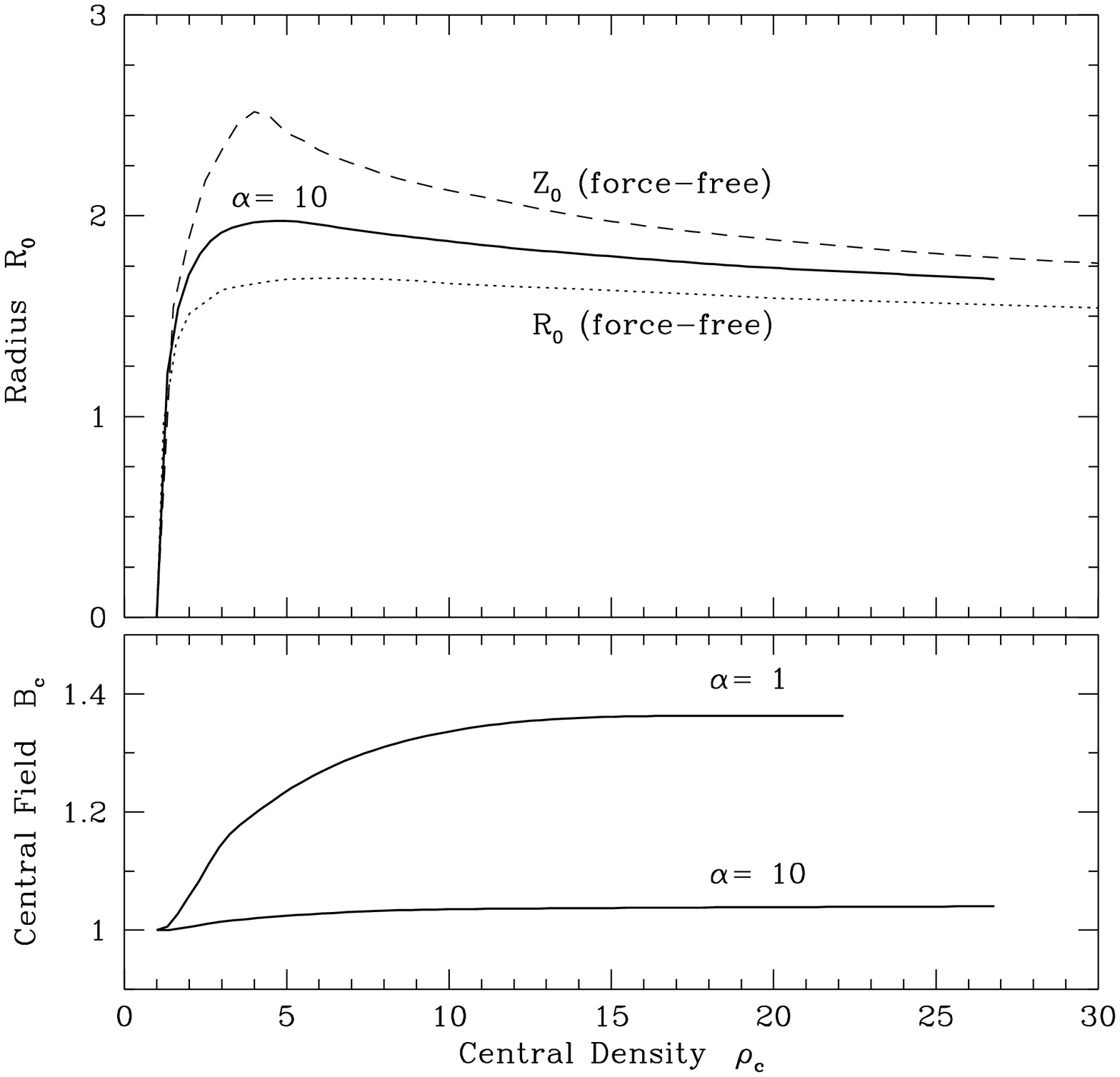}
\caption{{\it (top)} Radius versus density contrast for spheres under 
the tidal boundary condition with $\alpha = 10$ (solid curves).  The 
results are nearly identical for $\alpha = 1$.  The equatorial 
and polar radii of the force-free sequence are also shown. 
{\it (bottom)} Central magnetic field versus density contrast, 
along the $\alpha = 1$ and $\alpha = 10$ sequences.}
\efig

\bfig
\plotone{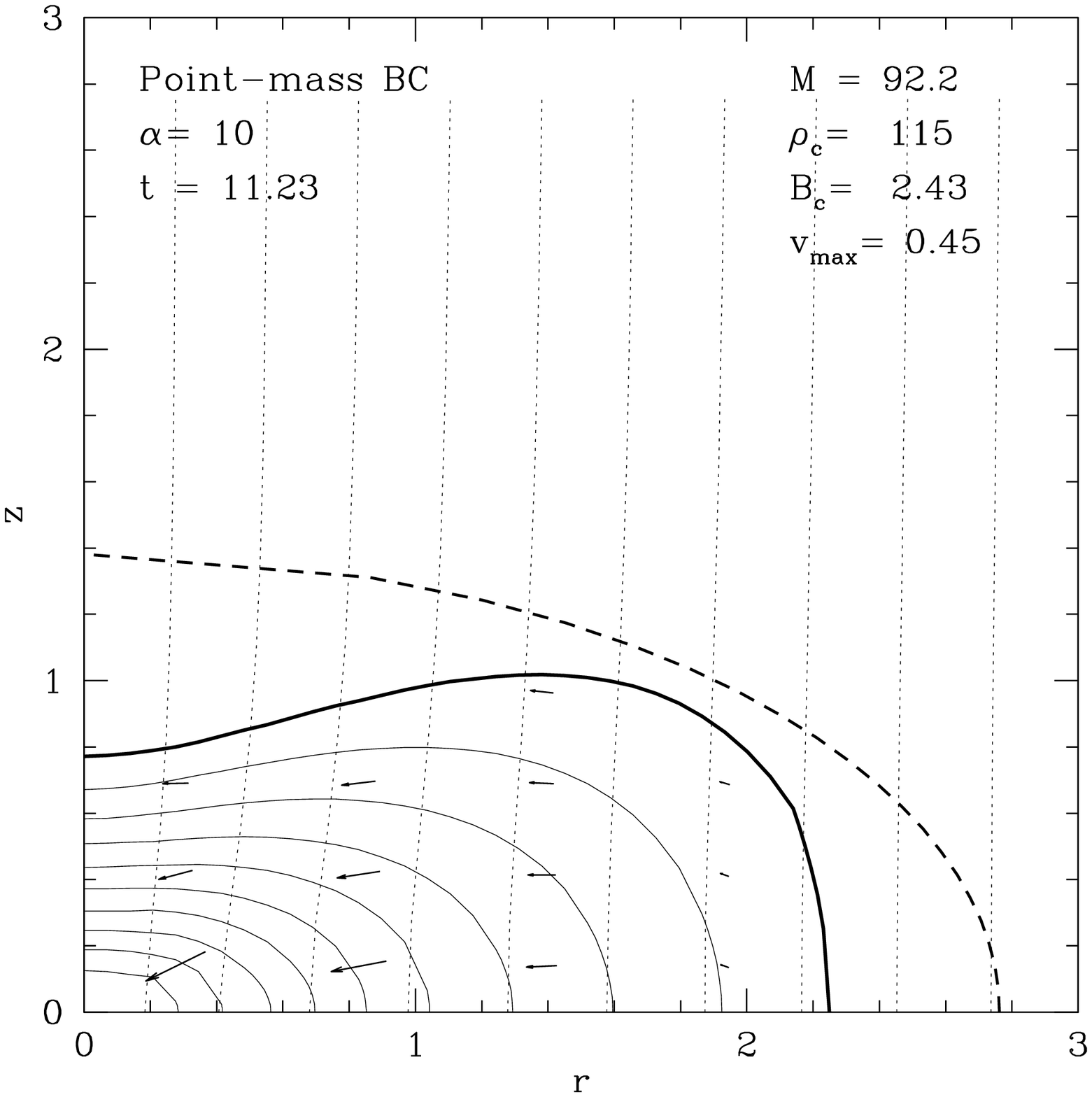}
\caption{
Final converged state of the isolated, 2:1 oblate cloud shown in 
Fig.\ \ref{fig-obinit}.  Arrows indicate neutral velocities, 
${\bf v}$.  The dashed curve shows the initial cloud boundary.}  
\efig

\bfig
\plotone{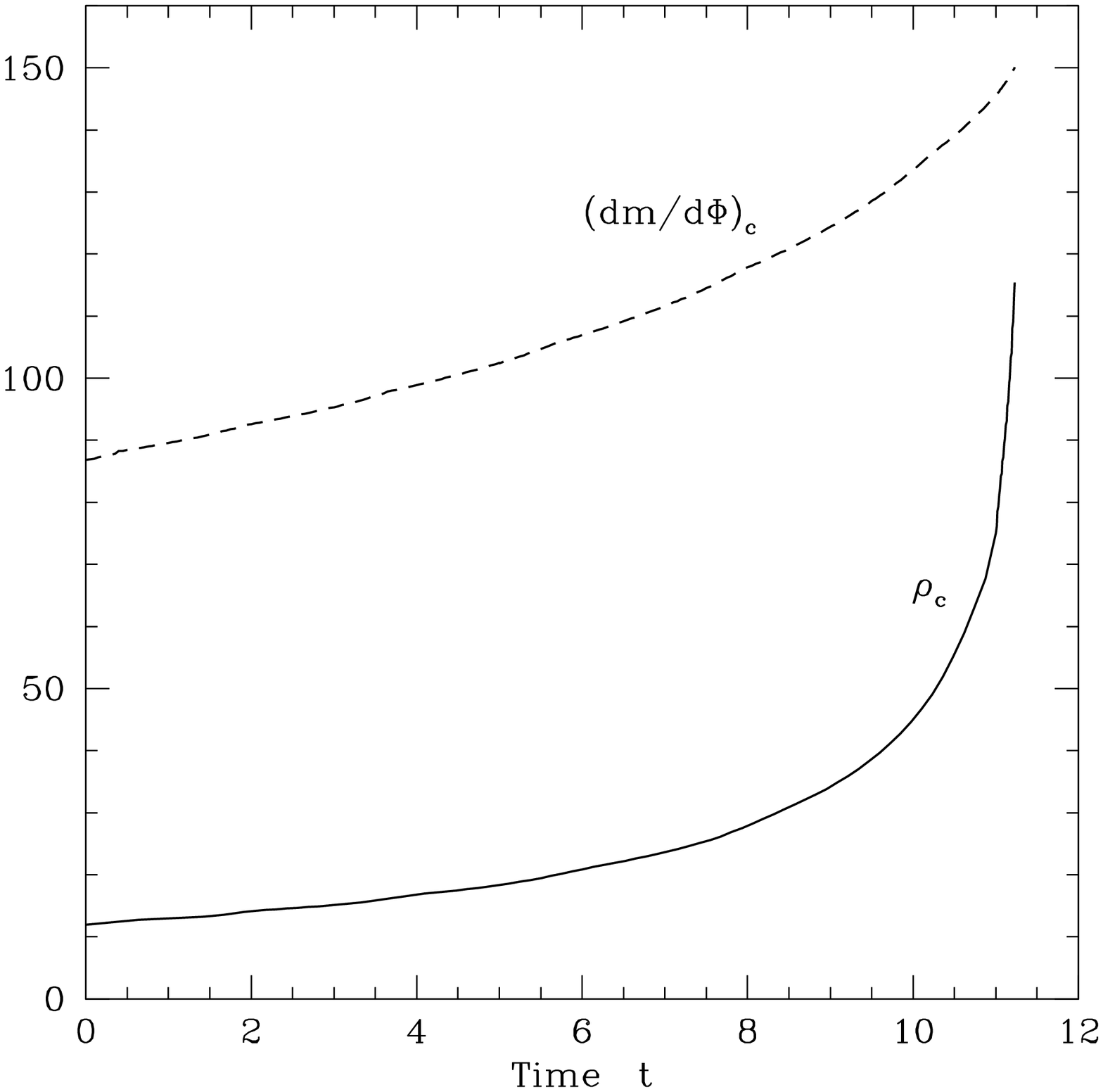}
\caption{
Time evolution of the central density (solid curve) and central 
mass-to-flux (dashed curve) for the oblate initial state of Fig.\ 
\ref{fig-obinit}.}
\efig

\bfig
\plotone{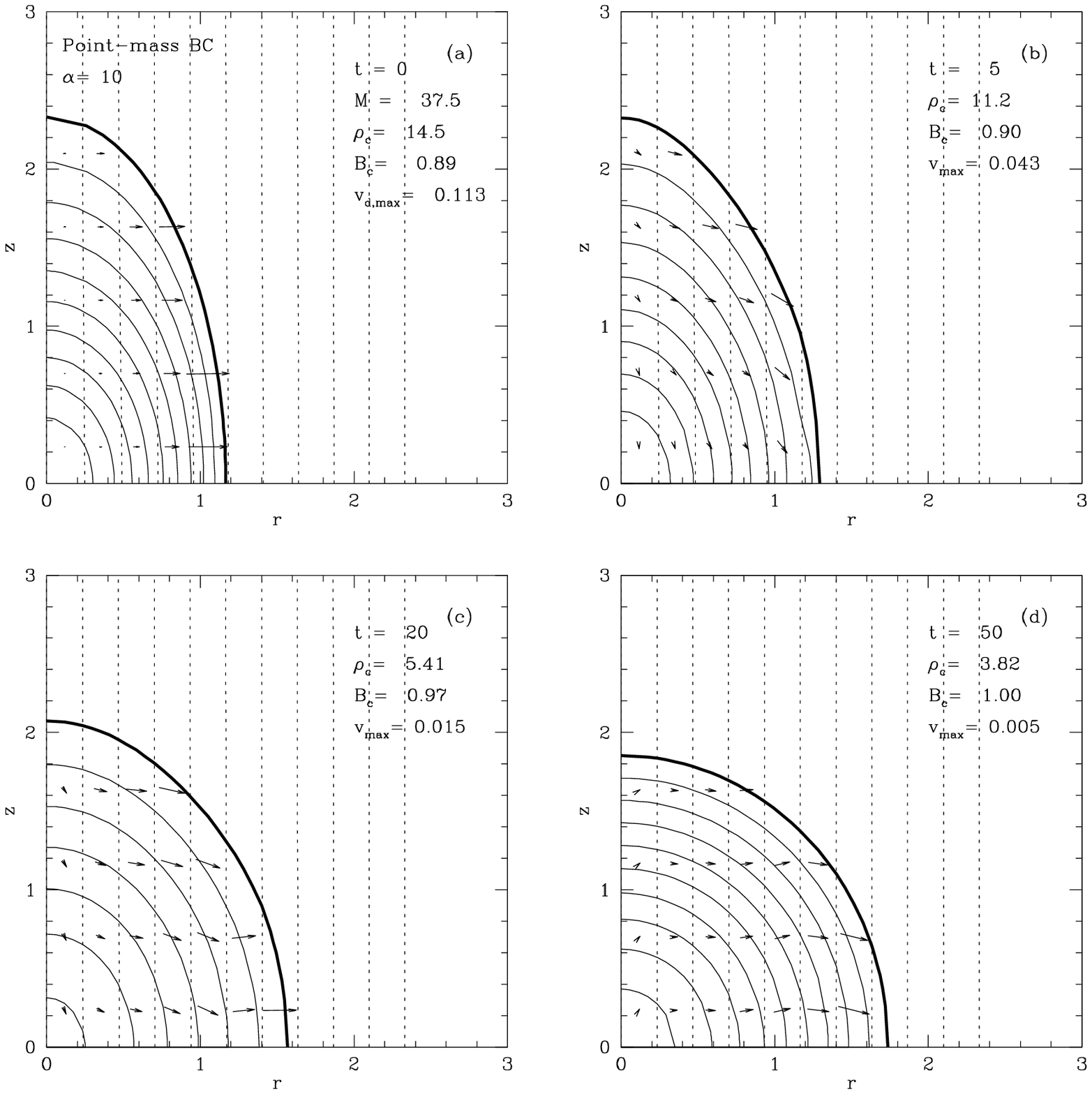}
\caption{
Time evolution of an isolated, 2:1 prolate cloud with $\alpha = 10$. 
Panel (a) shows the density contours (solid curves) and magnetic field 
lines (dotted curves) in the initial state.  Panels (b), (c), and (d) 
show the cloud at the times $t = 5, 20$, and 50, respectively.  
Arrows indicate drift velocities in panel (a), neutral velocities in 
panels (b), (c), and (d).
\label{fig-plevol}} 
\efig

\bfig
\plotone{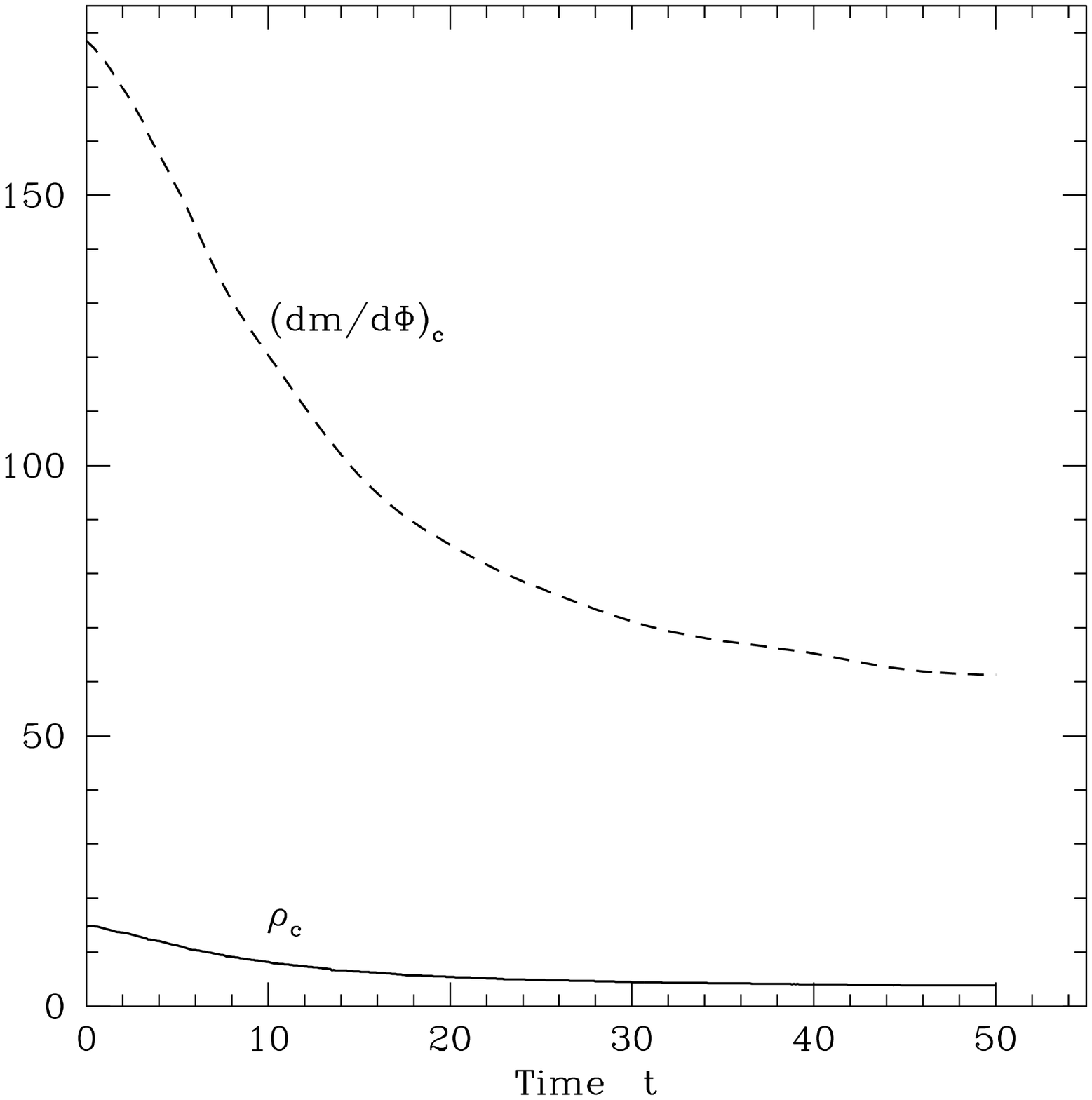}
\caption{
Time evolution of the central density (solid curve) and central 
mass-to-flux (dashed curve) for the initially prolate state of Fig.\ 
\ref{fig-plevol}.}
\efig

\bfig
\plotone{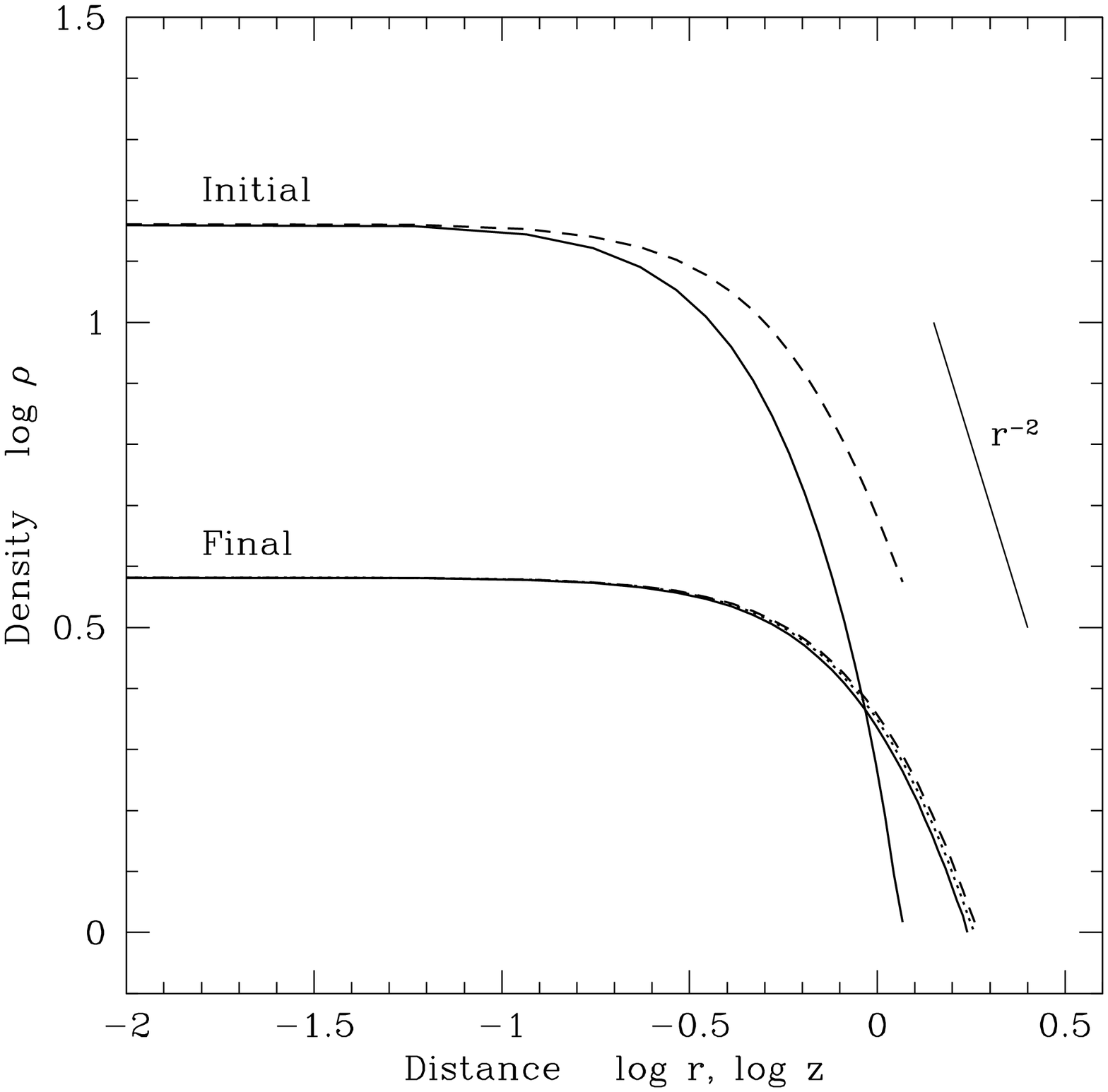}
\caption{
Density profiles of the prolate cloud in its initial and final 
states.  Solid curves show the density along the midplane $(z=0)$; 
dashed curves, along the pole $(r=0)$.  The density profile of an 
isolated, spherical cloud with the same central density is shown by 
a dotted curve.} 
\efig

\bfig
\plotone{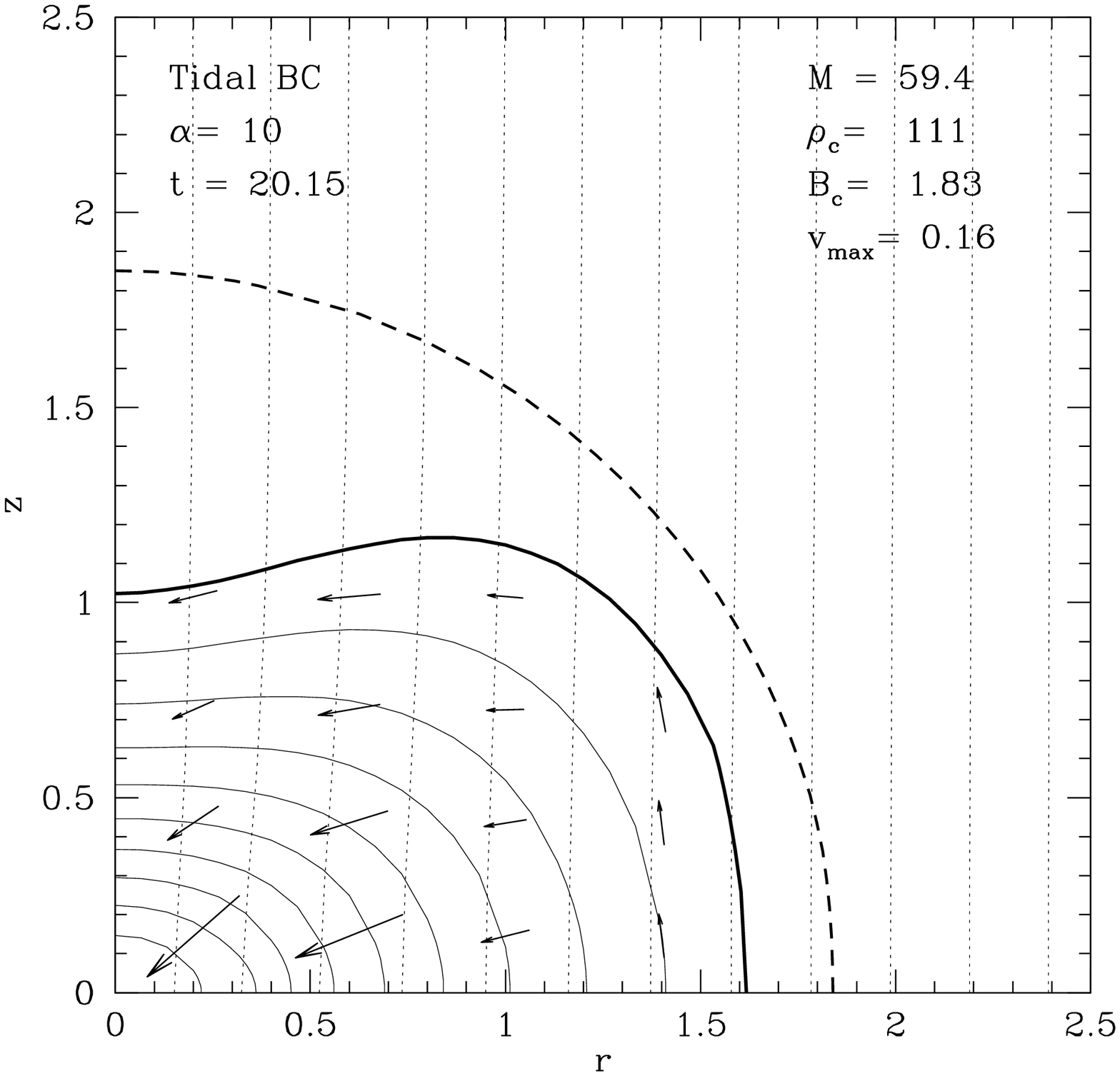}
\caption{
Final converged state of the tidally-stressed, spherical cloud 
shown in Fig.\ 10. Arrows indicate neutral velocities, 
${\bf v}$.  The dashed curve shows the initial cloud boundary.}  
\efig

\clearpage
\bfig
\plotone{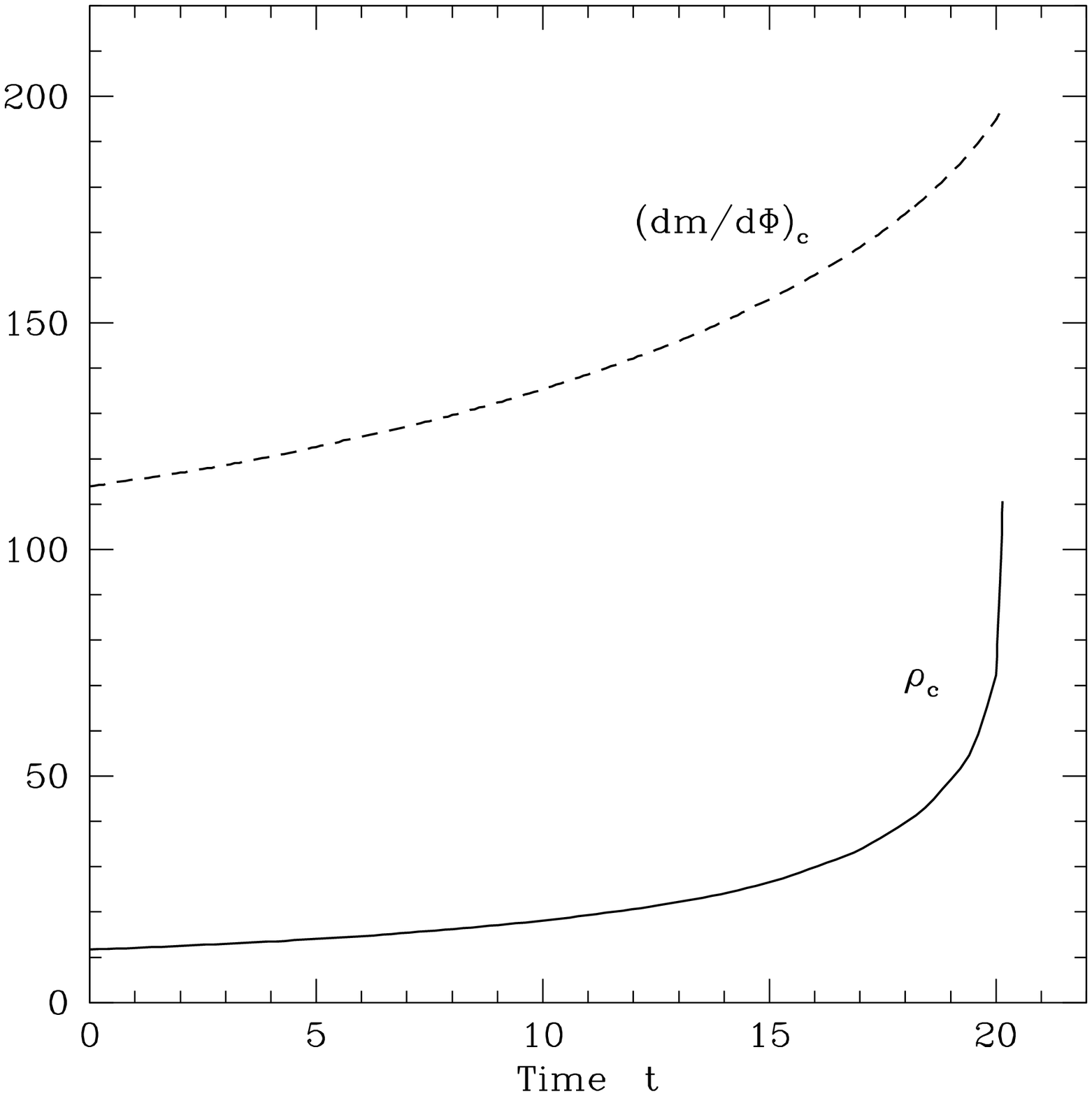}
\caption{
Time evolution of the central density (solid curve) and central 
mass-to-flux (dashed curve) for the spherical cloud of Fig.\ 
10.}
\efig

\bfig
\plotone{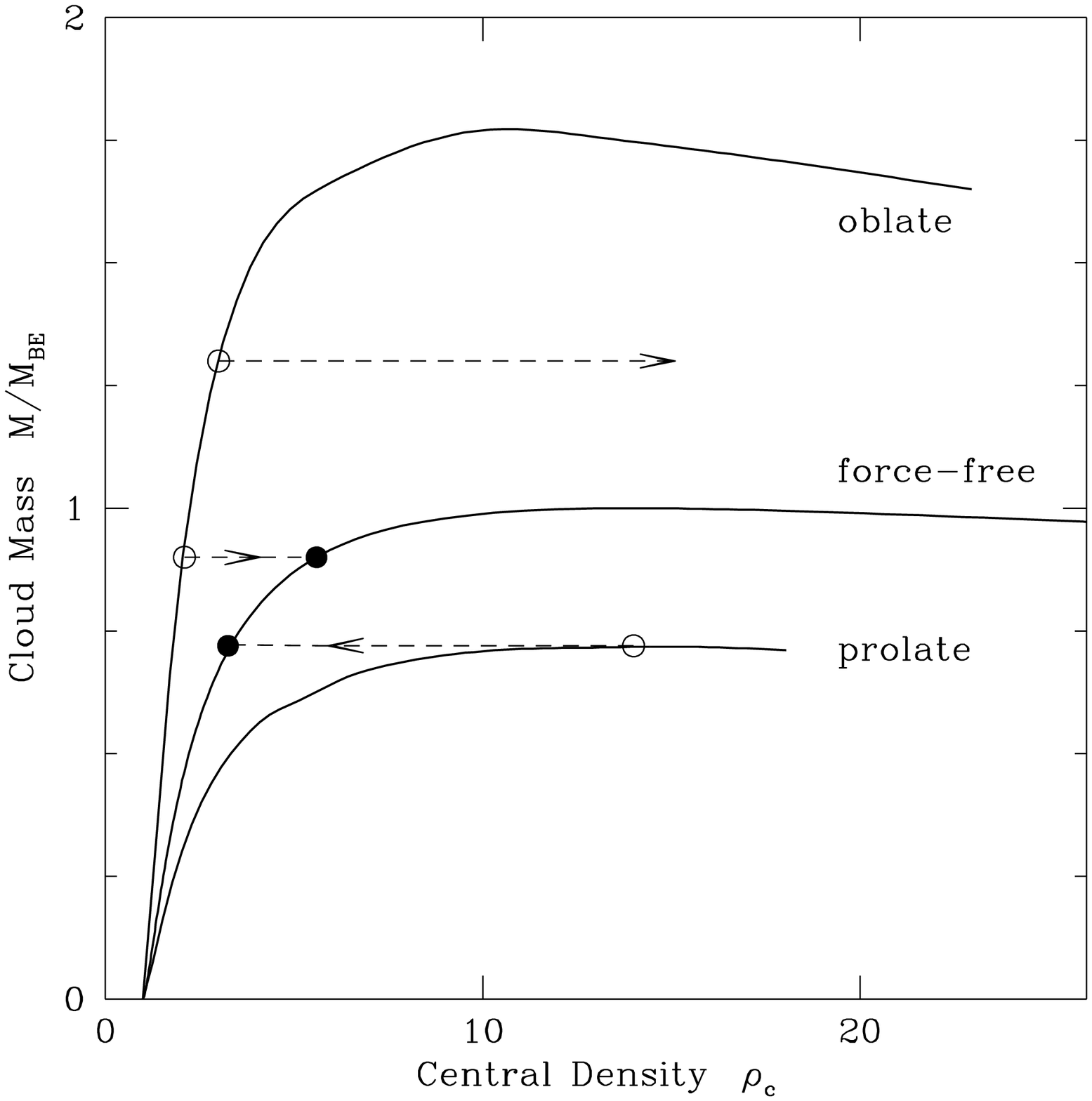}
\caption{
Mass as a function of density contrast for isolated equilibria 
considered in this paper (solid curves).  Results are displayed 
for $\alpha = 10$ only.  Initial states are represented by open 
circles, final states by filled circles.}  
\efig

\end{document}